\documentclass[aps, prb, reprint,superscriptaddress]{revtex4-2}

\usepackage{graphicx}
\usepackage{dcolumn}
\usepackage{bm}
\usepackage{amsmath,amssymb}
\usepackage{color} 
\usepackage{epstopdf}
\usepackage{soul}
\usepackage{siunitx}
\newcommand{\ignore}[1]{}

\begin{document}

\preprint{APS/123-QED}

\title{Noise properties of a Josephson parametric oscillator}

\author{Gopika Lakshmi Bhai}\email{gopika.lakshmibhai@gmail.com}
\affiliation{%
Graduate School of Science, Tokyo University of Science,1–3 Kagurazaka, Shinjuku, Tokyo 162–0825, Japan
}%
\affiliation{%
RIKEN Center for Quantum Computing (RQC), 2–1 Hirosawa, Wako, Saitama 351–0198, Japan
}%

\author{Hiroto Mukai}%
\affiliation{%
Graduate School of Science, Tokyo University of Science,1–3 Kagurazaka, Shinjuku, Tokyo 162–0825, Japan
}%
\affiliation{%
RIKEN Center for Quantum Computing (RQC), 2–1 Hirosawa, Wako, Saitama 351–0198, Japan
}%

\author{Tsuyoshi Yamamoto}

\affiliation{%
Research Institute for Science and Technology, Tokyo University of Science, 1-3 Kagurazaka, Shinjuku-ku, Tokyo 162-8601, Japan
}%

\affiliation{%
Secure System Platform Research Laboratories, NEC Corporation, Kawasaki, Kanagawa 211-0011, Japan
}%

\author{ Jaw-Shen Tsai}\email{tsai@riken.jp}
\affiliation{%
Graduate School of Science, Tokyo University of Science,1–3 Kagurazaka, Shinjuku, Tokyo 162–0825, Japan
}%

\affiliation{%
RIKEN Center for Quantum Computing (RQC), 2–1 Hirosawa, Wako, Saitama 351–0198, Japan
}%

\affiliation{%
Research Institute for Science and Technology, Tokyo University of Science, 1-3 Kagurazaka, Shinjuku-ku, Tokyo 162-8601, Japan
}%

\date{\today}


\begin{abstract}
We perform the noise spectroscopy of a Josephson parametric oscillator (JPO) by implementing a microwave homodyne interferometric measurement scheme. We observe the fluctuations in the self-oscillating output field of the JPO for a long 10 s time interval in a single shot measurement and characterize the phase and amplitude noise. Furthermore, we investigate the effects of the pump strength on the output noise power spectra of the JPO. We found strong fluctuations in the phase with a $1/f^2$ characteristics in the phase noise power spectrum, which is suppressed by increasing the pump strength.

\end{abstract}

\maketitle


\section{\label{sec:level1}Introduction}
 The study of noise in physical systems has a long history~\cite{kogan_1996}. The measurement and performance aspects of the physical systems have long been considered to be limited by the noise~\cite{johnson1928thermal,radeka196911overf,mcdermott2009materials,dutta1981low}. Various studies have been done extensively to understand the noise sources and the possibility of evading the noise~\cite{thermalnoise,suppressingchargenoise,greywall1994evading,shnirman2005lownoise,giovannetti2004quantum,clerk2010noise}. However, evading the intrinsic noise of a physical system is highly challenging and is a fundamental limitation~\cite{noiselimit_amp,goldberg1991theorylaser,caves1982quantum}. Recently, several advances have led to a renewed interest in the characterization and mitigation of noise in the field of circuit quantum electrodynamics (c-QED)~\cite{review_oliver,clerk2010noise,siddiqi2021qubit_noise,1/f_paladino20141} --- one of the most promising candidates in quantum computing and quantum information processing~\cite{kwon2021gate,gu2017review,devoret2005implementing}. The ongoing quest to build a quantum computer has intensified the efforts to study the noise in qubits~\cite{review_oliver,clerk2010noise,siddiqi2021qubit_noise} --- the basic building block of a quantum computer. However, the noise properties of other essential components used for the readout of the qubit, such as Josephson parametric amplifier (JPA) or JPO, are rarely explored. 
 
 A typical c-QED measurement network operates at frequencies of a few GHz and a temperature of around 10 mK. Quantum devices such as qubits are commonly read out at a few photon regimes to protect from measurement backaction~\cite{clerk2010noise, blais2021circuit,gambetta2006qubit}. Low noise amplification of the signal is a practical need to detect weak photons from the quantum devices operating in the microwave regime. JPA, a typical parametric device consisting of a superconducting resonator integrated with Josephson elements~\cite{yurke1988observationJPA,roy2018quantumJPA,eddins2019highJPA,yamamoto2008flux,aumentado2020superconducting} overcomes this obstacle by effectively amplifying the signal by adding a minimum noise allowed by the fundamental law of quantum mechanics~\cite{caves1982quantum,haus1962quantum}. These devices have become an essential component of the readout chain since they attain the quantum-limited amplification by high-frequency modulation of the inductance of the nonlinear Josephson element~\cite{yamamoto2008flux,sandberg2008tuning,vijay2010phase}. When the modulation amplitude exceeds the instability threshold, self-sustained oscillations start to build up~
 \cite{nayfeh2008nonlinear}, and it works as an oscillator --- Josephson parametric oscillator~\cite{lin2014josephson,krantz2013investigation,bengtsson2018nondegenerateJPO}. Due to the new accessible parameter regimes as a consequence of the strong nonlinear properties of JPO, notable studies have been done demonstrating the generation of squeezed states~\cite{squeezing_meaney2014quantum,squeezing2013parametric}, two-mode entanglement~\cite{squeezing2013parametric}, cat state engineering~\cite{goto2016bifurcation,goto2016universal,puri2017engineering,catqubit2014dynamically}, high-fidelity qubit readout~\cite{krantz2013investigation,lin2014josephson,singleshot_krantz}, etc. Nonetheless, the study of the noise properties of the JPO is left unaddressed. 
 
JPOs, like any other oscillators, have ubiquitous noise properties, which gives rise to a finite oscillation linewidth typically ranging from a few kHz to Hz depending on the operating parameters ~\cite{Casimir,bengtsson2018nondegenerateJPO,yurke_noise1991behavior,JPO_Yurke,olsson1988lowJPO}. Several unifying theories explain the noise characteristics of an oscillator~\cite{uni_hajimiri1998general,demir2000phaseunifying,uni_sauvage1977phase}. Over the last few decades, various experimental and theoretical studies in the optical domain have investigated the noise properties of parametric oscillators and lasers extensively~\cite{opo_reynaud1987quantum,imp_opo_1991phase,laser_henry1983theory,opo_debuisschert1989observation,dykman2015critical}. Noise spectroscopic studies in the optics field show the limiting factor of the finite linewidth of lasers~\cite{goldberg1991theorylaser,hinkley1969directlaser} described by the celebrated work of Schawlow-Townes~\cite{henry1983theoryST}. These theories and experimental observations shed light on investigating the noise properties of the JPO in superconducting circuit systems, where a detailed study of the noise properties of the JPO is yet to be investigated.

In this work, we present the experimental study of the noise characterization of a JPO pumped above its parametric threshold, where the phase coherence of the output photons from the oscillator is investigated. We perform the spectral analysis of noise in the phase and amplitude quadrature at low frequencies and explore the possible noise sources in JPO.

\section{Experimental setup}
\begin{figure}
\begin{center}
\includegraphics[keepaspectratio]{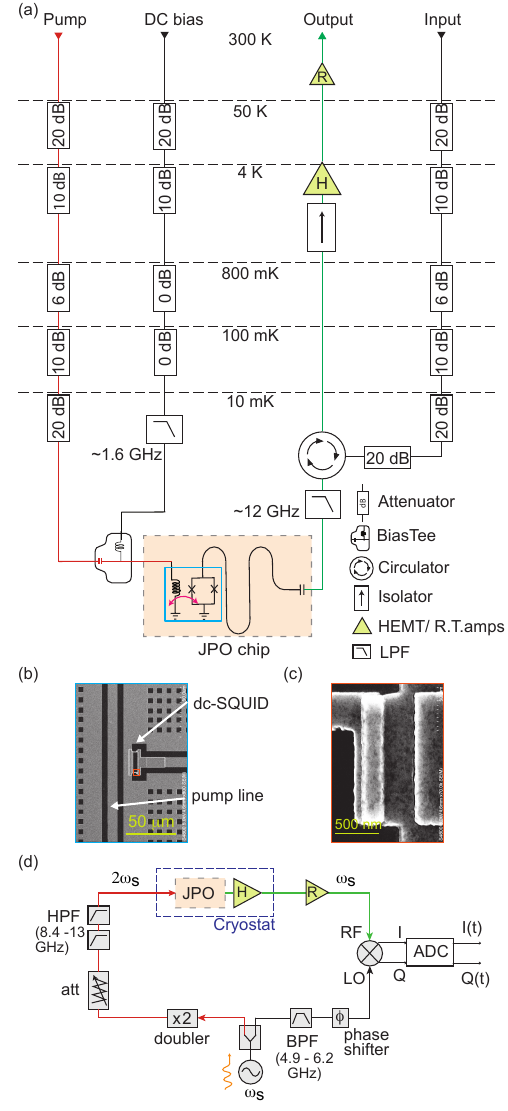} 
\end{center}
\caption{\label{Fig1} 
(a)~Schematic illustration of the cryogenic microwave measurement setup.
(b)~Scanning electron microscope (SEM) image showing SQUID consisting of two parallelly connected Josephson junctions. The SQUID made of aluminum is galvanically coupled to the central conductor of the CPW patterned on a niobium film which is deposited on a silicon substrate through sputtering. There are ground plane holes to pin the trapped flux on the chip.  (c)~SEM image showing Josephson junction made of Al/AlOx/Al. (d)~Microwave interferometric setup to measure the phase noise of the JPO: input lines are color-coded with the cryogenic circuit lines in Fig.~\ref{Fig1}(a), which describes the signal flow.}
\end{figure}

Our device consists of a $\lambda/4$ resonator made with a segment of coplanar waveguide (CPW) terminated by a dc superconducting quantum interference device (SQUID)~\cite{yamamoto2008flux}. The CPW resonator is fabricated by etching out the sputtered niobium on a silicon wafer~\cite{frunzio2005fabrication}. The Josephson junctions are then made by  standard double-angle shadow evaporation of aluminum [Fig.~\ref{Fig1}(b, c)]. The presence of the SQUID makes the resonance frequency of the resonator tunable by a variable Josephson inductance $L_{\mathrm{J}}$ which follows a relation $L_{\mathrm{J}}=\Phi_{0}/(4 \pi I_{\mathrm{c}}\left|\cos \left(\pi \Phi_{\mathrm{dc}} / \Phi_{0}\right)\right|)$. Here, $I_{\mathrm{c}}$ is the critical current through the junction, $\Phi_{\mathrm{dc}}$ is the dc magnetic flux through the SQUID,  and $\Phi_{0}$ is the flux quantum. By changing the flux through the dc-SQUID $\Phi_{\rm{dc}}$, the Josephson inductance can be varied. 

An input signal and a pump signal are applied to the device through the attenuated lines with filters to the respective ports on the chip. On-chip dc bias is applied through the same pump line using a bias-tee. The measurements of the device are carried out in a cryogenic environment using a dilution refrigerator at the base temperature of 10 mK [Fig.~\ref{Fig1}(a)]. The output signal from the device is routed through a microwave filter, circulator, and isolator, which is amplified using a high-electron-mobility transistor (HEMT) at 4K. The output signal is then further amplified at room temperature using R.T amplifiers.

At first, we characterize the system dynamics by scanning the resonator frequency dependence on the applied magnetic flux using a dc source. A weak probe signal is coupled to the cavity through a circulator in the mixing chamber, and the output from the device is analyzed using a vector network analyzer (VNA) by measuring the complex reflection coefficient $S_{11}$. The bare resonator frequency in the absence of flux was found to be $\omega_{\rm{r0}}/2\pi = 6.239~\si{\giga\hertz}$. The estimated critical current for each Josephson junction is $1.89~\si{\micro\ampere}$ which is obtained from fitting the resonator spectrum. The external and internal cavity loss are found to be $\kappa_{\mathrm{ext}}/2 \pi =$ 11 MHz and $\kappa_{\mathrm{int}}/2 \pi =$ 0.3 MHz at a flux bias of $\Phi_{\mathrm{dc}}/\Phi_{0}=$ 0.35 with a resonance frequency of $\omega_{\mathrm{r}}/2\pi = 5.94~\si{\giga\hertz}$. To further characterize the operational properties of the JPA, we apply a pump signal with a frequency $\omega_{\mathrm{p}}\approx2\omega_{\mathrm{r}}$ where $\omega_{\mathrm{r}}$ is the resonator frequency at a specific flux point. The pump photons interact with the incoming signal photons and generate an amplified signal with an approximate gain of $20~\si{\decibel}$ at pump power, $P_{\rm{p}} = -63.4~\si{\decibel\milli{}}$.

\begin{figure}
\begin{center}
\includegraphics[keepaspectratio]{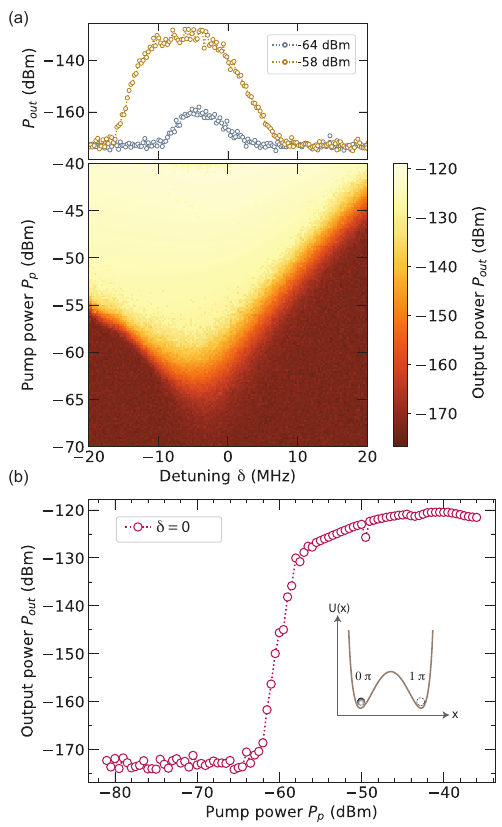} 
\end{center}
\caption{\label{Fig2} 
(a)~Span of the parametric regime of the JPO: The output power from the JPO (in the absence of a probe signal) is plotted against the pump power measured by changing the detuning $\delta$ between the resonator and pump frequencies. The top figure shows the cross-section at two fixed pump powers.
(b)~Output power from the JPO as a function of the applied pump power ($P_{\rm{p}}$). The static resonance frequency of the JPO is 5.94 GHz, at which the phase noise measurement of the JPO was carried out. The inset shows the qualitative picture of the bistable potential of the JPO, which has two stable minima corresponding to the oscillating states of $0\pi$ and $1\pi$~\cite{lin2014josephson}.}
\end{figure}

To investigate the photon generation, we apply the pump signal at twice the resonance frequency in the absence of the input signal and examine the output from the JPA by increasing the pump power $P_{\rm{p}}$. As the pump power increases beyond the parametric instability threshold, self-sustained oscillations build up exponentially in time inside the cavity. This can be understood as a second-order phase transition from a ground state below the threshold to an excited state above the  threshold~\cite{RevModPhys_equilibrium,wustmann2019_review_parametric}. In order to investigate the region of parametric instability, we choose a flux bias point on the flux-frequency curve at $\Phi_{\mathrm{dc}}/\Phi_{0}=$ 0.35, and we span the parametric plane by varying the pump-resonator detuning $\delta=|\omega_{\mathrm{r}}-\omega_{\mathrm{p}}/2|$, and pump power $P_{\rm{p}}$ as shown in Fig.~\ref{Fig2}(a). Here we see the regime of parametric oscillation throughout a well-defined interval of detuning $\delta$. The asymmetry of the parametric plane can be due to the pump-induced nonlinearity~\cite{krantz2013investigation}. Fig.~\ref{Fig2}(b) shows the output power from the JPO detected with a spectrum analyzer as a function of applied pump power $P_{\rm{p}}$ from a microwave source with $\omega_{\mathrm{p}}/2\pi =\,2\times5.94 ~\si{\giga\hertz}$. The power levels described are referred to the corresponding ports on the chip. As the pump power increases, crossing the threshold,  the output power from the JPO increases exponentially,  indicating the onset of parametric oscillation. In this above-threshold region, where the pump power is sufficiently strong, the nonlinearity of the system leads to bistability in the effective potential of the JPO field, as shown in the inset of Fig. 2(b). As a consequence of the bistability, the output oscillating field of the JPO has two stable states with a well-defined phase of either 0 or $\pi$~\cite{dykman2012fluctuating,lin2014josephson}.

Next, we measure the noise properties of the JPO operated at a pump frequency of $\omega_{\mathrm{p}}/2\pi =\,2\times5.94$ GHz at a fixed flux bias $\Phi_{\mathrm{dc}}/\Phi_{0}=0.35$. The measurement setup is shown in Fig.\ref{Fig1}(d). We use a homodyne interferometric measurement scheme where the output frequency of the JPO is mixed with a local oscillator at the same frequency to extract the phase and amplitude components of the signal using an IQ mixer. A microwave source at the frequency same as the JPO output signal frequency $\omega_{s}$ is split into two equal signals using a microwave splitter. One splited  signal is fed to the local oscillator of the IQ mixer through a band-pass filter and a phase shifter, whereas the other equally divided signal is connected to the JPO pump port by doubling the frequency using a frequency doubler. This pump signal is connected to a tunable attenuator to modulate the pump power. A series of carefully chosen high-pass and band-pass filters are connected to the pump and local oscillator signal path to filter out signals at unwanted frequencies. The transmitted output from the JPO is amplified with the HEMT and R.T amplifiers, which is then fed to the RF port of the IQ mixer. The I and Q signals from the IQ mixer are connected to the two channels of a 14-bit Keysight digitizer with an onboard field-programmable gate array (FPGA) for real-time sequencing. For the low-frequency phase noise measurements, the default sampling rate of the digitizer is reduced to 1MSa/s before channeling the signal to the data acquisition (DAQ). Using the custom FPGA image, this reduction in the sampling rate is made with 3-stage finite impulse response (FIR) filters with the decimation of 1/5, 1/4, and 1/25 in each step to obtain the desired sampling rate. The phase shifter in the local oscillator signal path is tuned to calibrate the offset between the I and Q signal due to the mixer imperfection and to utilize the digitizer's maximum dynamic range. We then observe this digitized data in real-time for a long period of 10 s in a single-shot. The I and Q signals are observed to be fluctuating about their mean. To quantify the JPO noise from this data, we first carry out the spectral noise analysis~\cite{gao2007noise,gao2011vacuum} of this digitized I and Q signal by studying the spectral domain noise covariance matrix $S(\nu)$ defined by,

\begin{align} 
\left\langle\delta \xi(\nu) \delta \xi^{\dagger}\left(\nu^{\prime}\right)\right\rangle &= S(\nu) \delta\left(\nu-\nu^{\prime}\right), \\
S(\nu) &=\left(\begin{array}{cc}S_{I I}(\nu) & S_{I Q}(\nu) \\ S_{I Q}^{*}(\nu) & S_{Q Q}(\nu)\end{array}\right).
\end{align}

Here, $\delta \xi(\nu)$ is the Fourier transform of the fluctuation in the I and Q signal about their mean, represented by $\delta \xi(t) = [\delta I(t),\delta Q(t)]^T$. Its Hermitian conjugate is given by $\delta \xi^{\dagger}(\nu)$. The diagonal terms in the matrix, $S_{I I}(\nu)$ and $S_{Q Q}(\nu)$ represent the auto power spectra. The off-diagonal element $S_{I Q}(\nu)$ represents the cross power spectra, and $S_{I Q}^{*}(\nu)$ represents its complex conjugate.

\begin{figure*}
\begin{center}
\includegraphics[keepaspectratio]{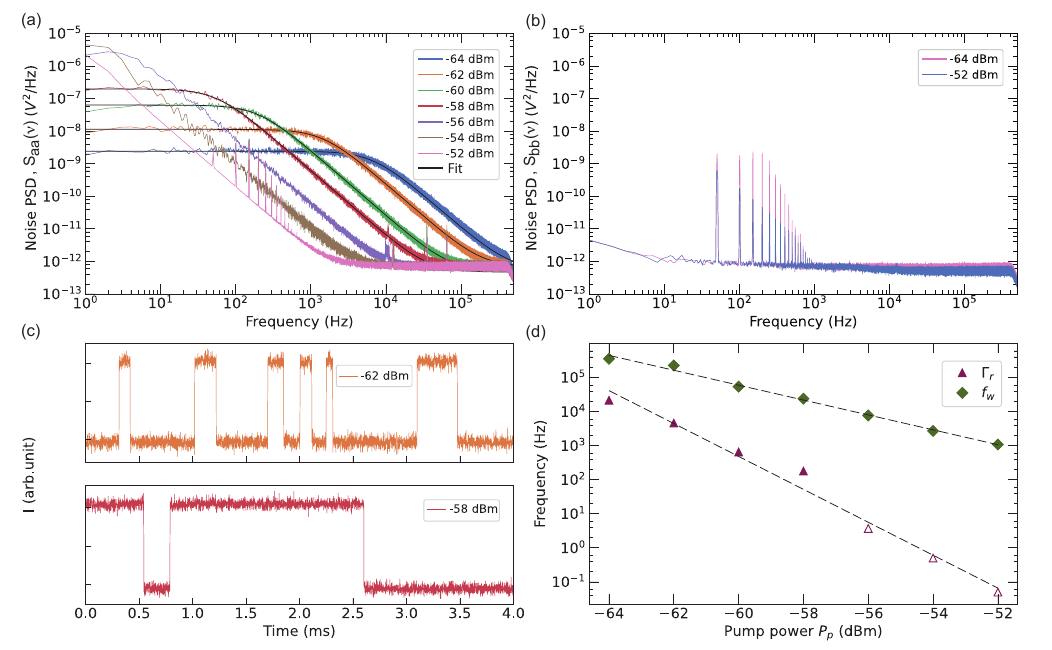}
\end{center}
\caption{\label{Fig3} 
(a)~Noise power spectral density (PSD) in the phase quadrature $S_{a a}(\nu)$ for different pump powers at a fixed pump frequency, $\omega_{\rm{p}}/2\pi =\,2\times5.94$ GHz. The solid black line shows the Lorentzian fit to the data. (b)~Noise PSD in the amplitude quadrature $S_{b b}(\nu)$ for two distinct pump powers. (c)~Trajectory of the I-quadrature as a function of time which is cropped from the full time-trace for two different pump powers. (d) Pump power dependence of corner frequencies, $\Gamma_{r}$ and $f_{w}$. Here, $\Gamma_{r}$ refers to $1/f^{2}$ roll-off and $f_{w}$ refers to the roll-off from $1/f^{2}$ to a white noise. $\Gamma_{r}$ shown in the shaded brown triangles are calculated from the Lorentzian fit, and those shown in the empty brown triangles are estimated from the time domain data by counting the number of switching events. Black dashed lines show the exponential fit.}
\end{figure*}

For a fixed pump power and frequency, the I and Q signals are recorded continuously for 10 s. This time trace data is then transferred to the CPU for further analysis. We calculate the spectral density using Welch's method ~\cite{welch1967use}. As described in Ref.~\citenum{gao2007noise}, we find that the imaginary part of $S_{I Q}$ is negligible, and the matrix can be diagonalized with an ordinary rotation $O(\nu)$ applied to $\rm{Re}\,S(\nu)$. Thus by diagonalizing it for all the frequencies, we obtain the eigenvalues as 

\begin{equation}
O^{T}(\nu) \operatorname{Re} S(\nu) O(\nu)=\left(\begin{array}{cc}S_{a a}(\nu) & 0 \\ 0 & S_{b b}(\nu)\end{array}\right).
\end{equation}

Here $S_{a a}$ and $S_{b b}$ represent the noise spectra in phase and amplitude quadrature. The measurements are repeated for different pump powers. The phase and amplitude power spectrum for each pump power is averaged five times. The HEMT noise floor and the measurement system noise floors are measured to ensure an adequate signal-to-noise ratio for our measurements.

\section{Results and discussions}

Fig.~\ref{Fig3}(a)  shows the phase noise power spectra of the JPO for different pump powers $P_{\rm{p}}$ ranging from $-64 ~\si{\decibel\milli{}}$ to $-52 ~\si{\decibel\milli{}}$, measured at a pump frequency of $\omega_{\mathrm{p}}/2\pi = 2\times 5.94~\si{\giga\hertz}$. At low frequencies, the phase noise spectra show a flat response, roll-off with a $1/f^{2}$ trend, and become white noise at higher frequencies. We note that the dip in the spectra at the highest frequency is due to the anti-aliasing filter. We can see that the phase noise is highly affected by the change in the pump power. When we gradually increase the pump power, we see suppression of phase noise in the $1/f^{2}$ characteristic region as the total noise envelope moves down to the lower frequencies.
On the contrary, the amplitude noise power spectra of the JPO in Fig.~\ref{Fig3}(b) are observed to be much lower than the phase noise spectra and are dominated by the HEMT noise floor except for a $1/f$ knee at lower frequencies contributed by the electronics. Within the range of our measurement sensitivity, amplitude noise shows no significant variation with the change in the pump power. The multiple peaks in the spectrum, from 50 Hz and its harmonics are an artifact of power grid poisoning~\cite{dugan1996electrical}, which could be eliminated using appropriate filters~\cite{filter_2014improving}.

Fig.~\ref{Fig3}(c) shows the trajectory of the JPO output in the I-quadrature for two different pump powers of $P_{\mathrm{p}} = -62 ~\si{\decibel\milli{}}$ and $P_{\mathrm{p}} = -58 ~\si{\decibel\milli{}}$, cropped from a 10 s long time trace recorded in a single shot measurement. We see that the signal is purely random, with discrete switching between two favorable states. It follows the typical characteristics of a random telegraphic noise which contribute to a Lorentzian noise spectrum in the frequency domain~\cite{RTS_lorentzian_noise}. In JPOs, the switching between two states originates due to the presence of a bistable potential whose minima correspond to $0\pi$ and $1\pi$ states~\cite{lin2014josephson}. We observe multiple flips between $0\pi$ and $1\pi$ states during an interval of 10 s.

For further analysis, we fit the noise spectra shown in Fig.~3(a) with a generalized Lorentzian~\cite{RTS_lorentzian_noise,feyyan_engineering} given by $S(f)=A\Gamma_{r}/(\pi^2 f^2 + \Gamma_{r}^2) + B$, where $\Gamma_{r}$, A and B are fitting parameters. $\Gamma_{r}$ is the corner frequency which is proportional to the interstate switching rate between $0\pi\rightarrow 1\pi$ and $1\pi\rightarrow 0\pi$ transitions~\cite{RTS_lorentzian_noise}. We also define a corner frequency $f_{w}$ at which the $1/f^{2}$ region rolls down to a white frequency noise. $f_{w}$ is evaluated from $1/f^{2}$ fit where it meets the white noise floor. These two corner frequencies for different pump powers are plotted in Fig.~\ref{Fig3}(d). For the lowest measured pump power, the switching rate is the highest. As the pump power gradually increases, the output oscillating field stays pinned to one of the two states for longer periods; as a result, the switching rate decreases. This can be understood by the theoretical description in Ref.~\citenum{lin2014josephson}. The barrier height of the bistable potential rises with the increase in the pump strength, which reduces the transition events between the two states, and the switching rate fall-off exponentially~\cite{friedman2003classical, razavy2013quantum_tunnelling}. We also note that the corner frequency $f_{w}$ decreases exponentially with the pump power. For $P_{\mathrm{p}} > -58 ~\si{\decibel\milli{}}$, extracting $\Gamma_{r}$  from the Lorentzian fit is non-viable since our sampling time interval is limited to 10 s. Hence, we count the number of switching events from the time domain data and estimate $\Gamma_{r}$. The empty triangles in Fig.~\ref{Fig3}(d) show the estimated $\Gamma_{r}$ for these powers. At these pump powers, the switching events are rare, and the oscillator field stays in one of its states for an extended period.

The switching between the bistable states could be a manifestation of a variety of mechanisms, including quantum fluctuation, thermal activation, or quantum activation~\cite{Casimir,dykman2015critical,dykman_quantum2007critical}. Further studies on the dynamic evolution of the oscillation output states of the JPO and the characteristic dependence of switching rate with the operating temperature would unravel the nature of this interstate transition~\cite{devoret_tunnelling_temperature}. More detailed theoretical studies on the switching rate specific to our nonlinear dynamical system and further experimental investigation of the dynamics of the JPO are the subject of our future work. Furthermore, the present system provides a platform to study and investigate the effects of a weak injection locking signal~\cite{markovic2019injection}, which is generally known to suppress the phase noise in self-sustained oscillators.

\section{Conclusion}
In conclusion, we have carried out the noise spectroscopy of the JPO with a microwave homodyne-interferometric measurement scheme. We discuss the analysis method to separate the noise power spectrum in the phase and amplitude quadrature. The extracted amplitude noise power spectra are much below the phase noise power spectra. The phase noise power spectra show signatures of random telegraphic noise due to the interstate transition of the output oscillating field between the two stable states of the JPO. This random telegraphic noise contributes to a $1/f^{2}$ noise characteristic to the phase noise power spectra. We studied this behavior by increasing the pump power and observed a significant reduction in the switching rate and suppression in the phase noise. Our noise analysis gives deeper insights into the dynamics and the noise characteristics of the JPO.

\begin{acknowledgments}
We are grateful to  V.~Sudhir, J.~Gao, and M.~I.~Dykman for their thoughtful comments on this research. We acknowledge fruitful discussions with E.~Rubiola, P.~Patil, R.~Wang, S.~Shirai, Y.~Zhou,  S.~Kwon, K.~Koshino, F.~Yoshihara, and Y.~Urade. Thanks to K.~Nittoh for support in fabrication. We also thank K.~Kikuchi for the technical support from Keysight. This research work was supported in part by JST Moonshot R\&D (JPMJMS2067), JST CREST (Grant No. JPMJCR1676 and JPMJCR1775) and a project JPNP16007, commissioned by New Energy and Industrial Technology Development Organization (NEDO), Japan.
\end{acknowledgments}

\bibliography{JPO_main.bib}

\begin{thebibliography}{71}%
\makeatletter
\providecommand \@ifxundefined [1]{%
 \@ifx{#1\undefined}
}%
\providecommand \@ifnum [1]{%
 \ifnum #1\expandafter \@firstoftwo
 \else \expandafter \@secondoftwo
 \fi
}%
\providecommand \@ifx [1]{%
 \ifx #1\expandafter \@firstoftwo
 \else \expandafter \@secondoftwo
 \fi
}%
\providecommand \natexlab [1]{#1}%
\providecommand \enquote  [1]{``#1''}%
\providecommand \bibnamefont  [1]{#1}%
\providecommand \bibfnamefont [1]{#1}%
\providecommand \citenamefont [1]{#1}%
\providecommand \href@noop [0]{\@secondoftwo}%
\providecommand \href [0]{\begingroup \@sanitize@url \@href}%
\providecommand \@href[1]{\@@startlink{#1}\@@href}%
\providecommand \@@href[1]{\endgroup#1\@@endlink}%
\providecommand \@sanitize@url [0]{\catcode `\\12\catcode `\$12\catcode
  `\&12\catcode `\#12\catcode `\^12\catcode `\_12\catcode `\%12\relax}%
\providecommand \@@startlink[1]{}%
\providecommand \@@endlink[0]{}%
\providecommand \url  [0]{\begingroup\@sanitize@url \@url }%
\providecommand \@url [1]{\endgroup\@href {#1}{\urlprefix }}%
\providecommand \urlprefix  [0]{URL }%
\providecommand \Eprint [0]{\href }%
\providecommand \doibase [0]{https://doi.org/}%
\providecommand \selectlanguage [0]{\@gobble}%
\providecommand \bibinfo  [0]{\@secondoftwo}%
\providecommand \bibfield  [0]{\@secondoftwo}%
\providecommand \translation [1]{[#1]}%
\providecommand \BibitemOpen [0]{}%
\providecommand \bibitemStop [0]{}%
\providecommand \bibitemNoStop [0]{.\EOS\space}%
\providecommand \EOS [0]{\spacefactor3000\relax}%
\providecommand \BibitemShut  [1]{\csname bibitem#1\endcsname}%
\let\auto@bib@innerbib\@empty
\bibitem [{\citenamefont {Kogan}(1996)}]{kogan_1996}%
  \BibitemOpen
  \bibfield  {author} {\bibinfo {author} {\bibfnamefont {S.}~\bibnamefont
  {Kogan}},\ }\href {https://doi.org/10.1017/CBO9780511551666} {\emph {\bibinfo
  {title} {Electronic Noise and Fluctuations in Solids}}}\ (\bibinfo
  {publisher} {Cambridge University Press, Cambridge, UK},\ \bibinfo {year}
  {1996})\BibitemShut {NoStop}%
\bibitem [{\citenamefont {Johnson}(1928)}]{johnson1928thermal}%
  \BibitemOpen
  \bibfield  {author} {\bibinfo {author} {\bibfnamefont {J.~B.}\ \bibnamefont
  {Johnson}},\ }\bibfield  {title} {\bibinfo {title} {Thermal agitation of
  electricity in conductors},\ }\href@noop {} {\bibfield  {journal} {\bibinfo
  {journal} {Phys. Rev.}\ }\textbf {\bibinfo {volume} {32}},\ \bibinfo {pages}
  {97} (\bibinfo {year} {1928})}\BibitemShut {NoStop}%
\bibitem [{\citenamefont {Radeka}(1969)}]{radeka196911overf}%
  \BibitemOpen
  \bibfield  {author} {\bibinfo {author} {\bibfnamefont {V.}~\bibnamefont
  {Radeka}},\ }\bibfield  {title} {\bibinfo {title} {$1/|f|$ noise in physical
  measurements},\ }\href@noop {} {\bibfield  {journal} {\bibinfo  {journal}
  {IEEE Trans. Nucl. Sci.}\ }\textbf {\bibinfo {volume} {16}},\ \bibinfo
  {pages} {17} (\bibinfo {year} {1969})}\BibitemShut {NoStop}%
\bibitem [{\citenamefont {McDermott}(2009)}]{mcdermott2009materials}%
  \BibitemOpen
  \bibfield  {author} {\bibinfo {author} {\bibfnamefont {R.}~\bibnamefont
  {McDermott}},\ }\bibfield  {title} {\bibinfo {title} {Materials origins of
  decoherence in superconducting qubits},\ }\href@noop {} {\bibfield  {journal}
  {\bibinfo  {journal} {IEEE Trans. Appl. Supercond.}\ }\textbf {\bibinfo
  {volume} {19}},\ \bibinfo {pages} {2} (\bibinfo {year} {2009})}\BibitemShut
  {NoStop}%
\bibitem [{\citenamefont {Dutta}\ and\ \citenamefont
  {Horn}(1981)}]{dutta1981low}%
  \BibitemOpen
  \bibfield  {author} {\bibinfo {author} {\bibfnamefont {P.}~\bibnamefont
  {Dutta}}\ and\ \bibinfo {author} {\bibfnamefont {P.}~\bibnamefont {Horn}},\
  }\bibfield  {title} {\bibinfo {title} {Low-frequency fluctuations in solids:
  $\frac{1}{f}$ noise},\ }\href@noop {} {\bibfield  {journal} {\bibinfo
  {journal} {Rev. Mod. Phys.}\ }\textbf {\bibinfo {volume} {53}},\ \bibinfo
  {pages} {497} (\bibinfo {year} {1981})}\BibitemShut {NoStop}%
\bibitem [{\citenamefont {Ivanov}\ and\ \citenamefont
  {Tobar}(2002)}]{thermalnoise}%
  \BibitemOpen
  \bibfield  {author} {\bibinfo {author} {\bibfnamefont {E.}~\bibnamefont
  {Ivanov}}\ and\ \bibinfo {author} {\bibfnamefont {M.}~\bibnamefont {Tobar}},\
  }\bibfield  {title} {\bibinfo {title} {``$\mathrm{R}$eal time" noise
  measurements with sensitivity exceeding the standard thermal noise limit},\
  }\href {https://doi.org/10.1109/TUFFC.2002.1026028} {\bibfield  {journal}
  {\bibinfo  {journal} {IEEE Trans. Ultrason. Ferroelectr. Freq. Control}\
  }\textbf {\bibinfo {volume} {49}},\ \bibinfo {pages} {1160} (\bibinfo {year}
  {2002})}\BibitemShut {NoStop}%
\bibitem [{\citenamefont {Schreier}\ \emph {et~al.}(2008)\citenamefont
  {Schreier}, \citenamefont {Houck}, \citenamefont {Koch}, \citenamefont
  {Schuster}, \citenamefont {Johnson}, \citenamefont {Chow}, \citenamefont
  {Gambetta}, \citenamefont {Majer}, \citenamefont {Frunzio}, \citenamefont
  {Devoret}, \citenamefont {Girvin},\ and\ \citenamefont
  {Schoelkopf}}]{suppressingchargenoise}%
  \BibitemOpen
  \bibfield  {author} {\bibinfo {author} {\bibfnamefont {J.~A.}\ \bibnamefont
  {Schreier}}, \bibinfo {author} {\bibfnamefont {A.~A.}\ \bibnamefont {Houck}},
  \bibinfo {author} {\bibfnamefont {J.}~\bibnamefont {Koch}}, \bibinfo {author}
  {\bibfnamefont {D.~I.}\ \bibnamefont {Schuster}}, \bibinfo {author}
  {\bibfnamefont {B.~R.}\ \bibnamefont {Johnson}}, \bibinfo {author}
  {\bibfnamefont {J.~M.}\ \bibnamefont {Chow}}, \bibinfo {author}
  {\bibfnamefont {J.~M.}\ \bibnamefont {Gambetta}}, \bibinfo {author}
  {\bibfnamefont {J.}~\bibnamefont {Majer}}, \bibinfo {author} {\bibfnamefont
  {L.}~\bibnamefont {Frunzio}}, \bibinfo {author} {\bibfnamefont {M.~H.}\
  \bibnamefont {Devoret}}, \bibinfo {author} {\bibfnamefont {S.~M.}\
  \bibnamefont {Girvin}},\ and\ \bibinfo {author} {\bibfnamefont {R.~J.}\
  \bibnamefont {Schoelkopf}},\ }\bibfield  {title} {\bibinfo {title}
  {Suppressing charge noise decoherence in superconducting charge qubits},\
  }\href@noop {} {\bibfield  {journal} {\bibinfo  {journal} {Phys. Rev. B}\
  }\textbf {\bibinfo {volume} {77}},\ \bibinfo {pages} {180502} (\bibinfo
  {year} {2008})}\BibitemShut {NoStop}%
\bibitem [{\citenamefont {Greywall}\ \emph {et~al.}(1994)\citenamefont
  {Greywall}, \citenamefont {Yurke}, \citenamefont {Busch}, \citenamefont
  {Pargellis},\ and\ \citenamefont {Willett}}]{greywall1994evading}%
  \BibitemOpen
  \bibfield  {author} {\bibinfo {author} {\bibfnamefont {D.}~\bibnamefont
  {Greywall}}, \bibinfo {author} {\bibfnamefont {B.}~\bibnamefont {Yurke}},
  \bibinfo {author} {\bibfnamefont {P.}~\bibnamefont {Busch}}, \bibinfo
  {author} {\bibfnamefont {A.}~\bibnamefont {Pargellis}},\ and\ \bibinfo
  {author} {\bibfnamefont {R.}~\bibnamefont {Willett}},\ }\bibfield  {title}
  {\bibinfo {title} {Evading amplifier noise in nonlinear oscillators},\
  }\href@noop {} {\bibfield  {journal} {\bibinfo  {journal} {Phys. Rev. Lett.}\
  }\textbf {\bibinfo {volume} {72}},\ \bibinfo {pages} {2992} (\bibinfo {year}
  {1994})}\BibitemShut {NoStop}%
\bibitem [{\citenamefont {Shnirman}\ \emph {et~al.}(2005)\citenamefont
  {Shnirman}, \citenamefont {Sch{\"o}n}, \citenamefont {Martin},\ and\
  \citenamefont {Makhlin}}]{shnirman2005lownoise}%
  \BibitemOpen
  \bibfield  {author} {\bibinfo {author} {\bibfnamefont {A.}~\bibnamefont
  {Shnirman}}, \bibinfo {author} {\bibfnamefont {G.}~\bibnamefont {Sch{\"o}n}},
  \bibinfo {author} {\bibfnamefont {I.}~\bibnamefont {Martin}},\ and\ \bibinfo
  {author} {\bibfnamefont {Y.}~\bibnamefont {Makhlin}},\ }\bibfield  {title}
  {\bibinfo {title} {Low-and high-frequency noise from coherent two-level
  systems},\ }\href@noop {} {\bibfield  {journal} {\bibinfo  {journal} {Phys.
  Rev. Lett.}\ }\textbf {\bibinfo {volume} {94}},\ \bibinfo {pages} {127002}
  (\bibinfo {year} {2005})}\BibitemShut {NoStop}%
\bibitem [{\citenamefont {Giovannetti}\ \emph {et~al.}(2004)\citenamefont
  {Giovannetti}, \citenamefont {Lloyd},\ and\ \citenamefont
  {Maccone}}]{giovannetti2004quantum}%
  \BibitemOpen
  \bibfield  {author} {\bibinfo {author} {\bibfnamefont {V.}~\bibnamefont
  {Giovannetti}}, \bibinfo {author} {\bibfnamefont {S.}~\bibnamefont {Lloyd}},\
  and\ \bibinfo {author} {\bibfnamefont {L.}~\bibnamefont {Maccone}},\
  }\bibfield  {title} {\bibinfo {title} {Quantum-enhanced measurements: beating
  the standard quantum limit},\ }\href@noop {} {\bibfield  {journal} {\bibinfo
  {journal} {Science}\ }\textbf {\bibinfo {volume} {306}},\ \bibinfo {pages}
  {1330} (\bibinfo {year} {2004})}\BibitemShut {NoStop}%
\bibitem [{\citenamefont {Clerk}\ \emph {et~al.}(2010)\citenamefont {Clerk},
  \citenamefont {Devoret}, \citenamefont {Girvin}, \citenamefont {Marquardt},\
  and\ \citenamefont {Schoelkopf}}]{clerk2010noise}%
  \BibitemOpen
  \bibfield  {author} {\bibinfo {author} {\bibfnamefont {A.~A.}\ \bibnamefont
  {Clerk}}, \bibinfo {author} {\bibfnamefont {M.~H.}\ \bibnamefont {Devoret}},
  \bibinfo {author} {\bibfnamefont {S.~M.}\ \bibnamefont {Girvin}}, \bibinfo
  {author} {\bibfnamefont {F.}~\bibnamefont {Marquardt}},\ and\ \bibinfo
  {author} {\bibfnamefont {R.~J.}\ \bibnamefont {Schoelkopf}},\ }\bibfield
  {title} {\bibinfo {title} {Introduction to quantum noise, measurement, and
  amplification},\ }\href@noop {} {\bibfield  {journal} {\bibinfo  {journal}
  {Rev. Mod. Phys.}\ }\textbf {\bibinfo {volume} {82}},\ \bibinfo {pages}
  {1155} (\bibinfo {year} {2010})}\BibitemShut {NoStop}%
\bibitem [{\citenamefont {Heffner}(1962)}]{noiselimit_amp}%
  \BibitemOpen
  \bibfield  {author} {\bibinfo {author} {\bibfnamefont {H.}~\bibnamefont
  {Heffner}},\ }\bibfield  {title} {\bibinfo {title} {The fundamental noise
  limit of linear amplifiers},\ }\href
  {https://doi.org/10.1109/JRPROC.1962.288130} {\bibfield  {journal} {\bibinfo
  {journal} {Proc. IRE 50}\ }\textbf {\bibinfo {volume} {50}},\ \bibinfo
  {pages} {1604} (\bibinfo {year} {1962})}\BibitemShut {NoStop}%
\bibitem [{\citenamefont {Goldberg}\ \emph {et~al.}(1991)\citenamefont
  {Goldberg}, \citenamefont {Milonni},\ and\ \citenamefont
  {Sundaram}}]{goldberg1991theorylaser}%
  \BibitemOpen
  \bibfield  {author} {\bibinfo {author} {\bibfnamefont {P.}~\bibnamefont
  {Goldberg}}, \bibinfo {author} {\bibfnamefont {P.~W.}\ \bibnamefont
  {Milonni}},\ and\ \bibinfo {author} {\bibfnamefont {B.}~\bibnamefont
  {Sundaram}},\ }\bibfield  {title} {\bibinfo {title} {Theory of the
  fundamental laser linewidth},\ }\href@noop {} {\bibfield  {journal} {\bibinfo
   {journal} {Phys. Rev. A}\ }\textbf {\bibinfo {volume} {44}},\ \bibinfo
  {pages} {1969} (\bibinfo {year} {1991})}\BibitemShut {NoStop}%
\bibitem [{\citenamefont {Caves}(1982)}]{caves1982quantum}%
  \BibitemOpen
  \bibfield  {author} {\bibinfo {author} {\bibfnamefont {C.~M.}\ \bibnamefont
  {Caves}},\ }\bibfield  {title} {\bibinfo {title} {Quantum limits on noise in
  linear amplifiers},\ }\href@noop {} {\bibfield  {journal} {\bibinfo
  {journal} {Phys. Rev. D}\ }\textbf {\bibinfo {volume} {26}},\ \bibinfo
  {pages} {1817} (\bibinfo {year} {1982})}\BibitemShut {NoStop}%
\bibitem [{\citenamefont {Krantz}\ \emph {et~al.}(2019)\citenamefont {Krantz},
  \citenamefont {Kjaergaard}, \citenamefont {Yan}, \citenamefont {Orlando},
  \citenamefont {Gustavsson},\ and\ \citenamefont {Oliver}}]{review_oliver}%
  \BibitemOpen
  \bibfield  {author} {\bibinfo {author} {\bibfnamefont {P.}~\bibnamefont
  {Krantz}}, \bibinfo {author} {\bibfnamefont {M.}~\bibnamefont {Kjaergaard}},
  \bibinfo {author} {\bibfnamefont {F.}~\bibnamefont {Yan}}, \bibinfo {author}
  {\bibfnamefont {T.~P.}\ \bibnamefont {Orlando}}, \bibinfo {author}
  {\bibfnamefont {S.}~\bibnamefont {Gustavsson}},\ and\ \bibinfo {author}
  {\bibfnamefont {W.~D.}\ \bibnamefont {Oliver}},\ }\bibfield  {title}
  {\bibinfo {title} {A quantum engineer's guide to superconducting qubits},\
  }\href@noop {} {\bibfield  {journal} {\bibinfo  {journal} {Appl. Phys. Rev.}\
  }\textbf {\bibinfo {volume} {6}},\ \bibinfo {pages} {021318} (\bibinfo {year}
  {2019})}\BibitemShut {NoStop}%
\bibitem [{\citenamefont {Siddiqi}(2021)}]{siddiqi2021qubit_noise}%
  \BibitemOpen
  \bibfield  {author} {\bibinfo {author} {\bibfnamefont {I.}~\bibnamefont
  {Siddiqi}},\ }\bibfield  {title} {\bibinfo {title} {Engineering
  high-coherence superconducting qubits},\ }\href@noop {} {\bibfield  {journal}
  {\bibinfo  {journal} {Nat. Rev. Mat.}\ }\textbf {\bibinfo {volume} {6}},\
  \bibinfo {pages} {875} (\bibinfo {year} {2021})}\BibitemShut {NoStop}%
\bibitem [{\citenamefont {Paladino}\ \emph {et~al.}(2014)\citenamefont
  {Paladino}, \citenamefont {Galperin}, \citenamefont {Falci},\ and\
  \citenamefont {Altshuler}}]{1/f_paladino20141}%
  \BibitemOpen
  \bibfield  {author} {\bibinfo {author} {\bibfnamefont {E.}~\bibnamefont
  {Paladino}}, \bibinfo {author} {\bibfnamefont {Y.}~\bibnamefont {Galperin}},
  \bibinfo {author} {\bibfnamefont {G.}~\bibnamefont {Falci}},\ and\ \bibinfo
  {author} {\bibfnamefont {B.}~\bibnamefont {Altshuler}},\ }\bibfield  {title}
  {\bibinfo {title} {1/f noise: Implications for solid-state quantum
  information},\ }\href@noop {} {\bibfield  {journal} {\bibinfo  {journal}
  {Rev. Mod. Phys.}\ }\textbf {\bibinfo {volume} {86}},\ \bibinfo {pages} {361}
  (\bibinfo {year} {2014})}\BibitemShut {NoStop}%
\bibitem [{\citenamefont {Kwon}\ \emph {et~al.}(2021)\citenamefont {Kwon},
  \citenamefont {Tomonaga}, \citenamefont {Lakshmi~Bhai}, \citenamefont
  {Devitt},\ and\ \citenamefont {Tsai}}]{kwon2021gate}%
  \BibitemOpen
  \bibfield  {author} {\bibinfo {author} {\bibfnamefont {S.}~\bibnamefont
  {Kwon}}, \bibinfo {author} {\bibfnamefont {A.}~\bibnamefont {Tomonaga}},
  \bibinfo {author} {\bibfnamefont {G.}~\bibnamefont {Lakshmi~Bhai}}, \bibinfo
  {author} {\bibfnamefont {S.~J.}\ \bibnamefont {Devitt}},\ and\ \bibinfo
  {author} {\bibfnamefont {J.-S.}\ \bibnamefont {Tsai}},\ }\bibfield  {title}
  {\bibinfo {title} {Gate-based superconducting quantum computing},\
  }\href@noop {} {\bibfield  {journal} {\bibinfo  {journal} {J. Appl. Phys.}\
  }\textbf {\bibinfo {volume} {129}},\ \bibinfo {pages} {041102} (\bibinfo
  {year} {2021})}\BibitemShut {NoStop}%
\bibitem [{\citenamefont {Gu}\ \emph {et~al.}(2017)\citenamefont {Gu},
  \citenamefont {Kockum}, \citenamefont {Miranowicz}, \citenamefont {Liu},\
  and\ \citenamefont {Nori}}]{gu2017review}%
  \BibitemOpen
  \bibfield  {author} {\bibinfo {author} {\bibfnamefont {X.}~\bibnamefont
  {Gu}}, \bibinfo {author} {\bibfnamefont {A.~F.}\ \bibnamefont {Kockum}},
  \bibinfo {author} {\bibfnamefont {A.}~\bibnamefont {Miranowicz}}, \bibinfo
  {author} {\bibfnamefont {Y.-x.}\ \bibnamefont {Liu}},\ and\ \bibinfo {author}
  {\bibfnamefont {F.}~\bibnamefont {Nori}},\ }\bibfield  {title} {\bibinfo
  {title} {Microwave photonics with superconducting quantum circuits},\
  }\href@noop {} {\bibfield  {journal} {\bibinfo  {journal} {Phys. Rep.}\
  }\textbf {\bibinfo {volume} {718}},\ \bibinfo {pages} {1} (\bibinfo {year}
  {2017})}\BibitemShut {NoStop}%
\bibitem [{\citenamefont {Devoret}\ and\ \citenamefont
  {Martinis}(2004)}]{devoret2005implementing}%
  \BibitemOpen
  \bibfield  {author} {\bibinfo {author} {\bibfnamefont {M.~H.}\ \bibnamefont
  {Devoret}}\ and\ \bibinfo {author} {\bibfnamefont {J.~M.}\ \bibnamefont
  {Martinis}},\ }\bibfield  {title} {\bibinfo {title} {Implementing qubits with
  superconducting integrated circuits},\ }\href@noop {} {\bibfield  {journal}
  {\bibinfo  {journal} {Quantum Inf. Process.}\ }\textbf {\bibinfo {volume}
  {3}},\ \bibinfo {pages} {163} (\bibinfo {year} {2004})}\BibitemShut {NoStop}%
\bibitem [{\citenamefont {Blais}\ \emph {et~al.}(2021)\citenamefont {Blais},
  \citenamefont {Grimsmo}, \citenamefont {Girvin},\ and\ \citenamefont
  {Wallraff}}]{blais2021circuit}%
  \BibitemOpen
  \bibfield  {author} {\bibinfo {author} {\bibfnamefont {A.}~\bibnamefont
  {Blais}}, \bibinfo {author} {\bibfnamefont {A.~L.}\ \bibnamefont {Grimsmo}},
  \bibinfo {author} {\bibfnamefont {S.~M.}\ \bibnamefont {Girvin}},\ and\
  \bibinfo {author} {\bibfnamefont {A.}~\bibnamefont {Wallraff}},\ }\bibfield
  {title} {\bibinfo {title} {Circuit quantum electrodynamics},\ }\href@noop {}
  {\bibfield  {journal} {\bibinfo  {journal} {Rev. Mod. Phys.}\ }\textbf
  {\bibinfo {volume} {93}},\ \bibinfo {pages} {025005} (\bibinfo {year}
  {2021})}\BibitemShut {NoStop}%
\bibitem [{\citenamefont {Gambetta}\ \emph {et~al.}(2006)\citenamefont
  {Gambetta}, \citenamefont {Blais}, \citenamefont {Schuster}, \citenamefont
  {Wallraff}, \citenamefont {Frunzio}, \citenamefont {Majer}, \citenamefont
  {Devoret}, \citenamefont {Girvin},\ and\ \citenamefont
  {Schoelkopf}}]{gambetta2006qubit}%
  \BibitemOpen
  \bibfield  {author} {\bibinfo {author} {\bibfnamefont {J.}~\bibnamefont
  {Gambetta}}, \bibinfo {author} {\bibfnamefont {A.}~\bibnamefont {Blais}},
  \bibinfo {author} {\bibfnamefont {D.~I.}\ \bibnamefont {Schuster}}, \bibinfo
  {author} {\bibfnamefont {A.}~\bibnamefont {Wallraff}}, \bibinfo {author}
  {\bibfnamefont {L.}~\bibnamefont {Frunzio}}, \bibinfo {author} {\bibfnamefont
  {J.}~\bibnamefont {Majer}}, \bibinfo {author} {\bibfnamefont {M.~H.}\
  \bibnamefont {Devoret}}, \bibinfo {author} {\bibfnamefont {S.~M.}\
  \bibnamefont {Girvin}},\ and\ \bibinfo {author} {\bibfnamefont {R.~J.}\
  \bibnamefont {Schoelkopf}},\ }\bibfield  {title} {\bibinfo {title}
  {Qubit-photon interactions in a cavity: Measurement-induced dephasing and
  number splitting},\ }\href@noop {} {\bibfield  {journal} {\bibinfo  {journal}
  {Phys. Rev. A}\ }\textbf {\bibinfo {volume} {74}},\ \bibinfo {pages} {042318}
  (\bibinfo {year} {2006})}\BibitemShut {NoStop}%
\bibitem [{\citenamefont {Yurke}\ \emph {et~al.}(1988)\citenamefont {Yurke},
  \citenamefont {Kaminsky}, \citenamefont {Miller}, \citenamefont {Whittaker},
  \citenamefont {Smith}, \citenamefont {Silver},\ and\ \citenamefont
  {Simon}}]{yurke1988observationJPA}%
  \BibitemOpen
  \bibfield  {author} {\bibinfo {author} {\bibfnamefont {B.}~\bibnamefont
  {Yurke}}, \bibinfo {author} {\bibfnamefont {P.}~\bibnamefont {Kaminsky}},
  \bibinfo {author} {\bibfnamefont {R.}~\bibnamefont {Miller}}, \bibinfo
  {author} {\bibfnamefont {E.}~\bibnamefont {Whittaker}}, \bibinfo {author}
  {\bibfnamefont {A.}~\bibnamefont {Smith}}, \bibinfo {author} {\bibfnamefont
  {A.}~\bibnamefont {Silver}},\ and\ \bibinfo {author} {\bibfnamefont
  {R.}~\bibnamefont {Simon}},\ }\bibfield  {title} {\bibinfo {title}
  {Observation of 4.2-k equilibrium-noise squeezing via a josephson-parametric
  amplifier},\ }\href@noop {} {\bibfield  {journal} {\bibinfo  {journal} {Phys.
  Rev. Lett.}\ }\textbf {\bibinfo {volume} {60}},\ \bibinfo {pages} {764}
  (\bibinfo {year} {1988})}\BibitemShut {NoStop}%
\bibitem [{\citenamefont {Roy}\ and\ \citenamefont
  {Devoret}(2018)}]{roy2018quantumJPA}%
  \BibitemOpen
  \bibfield  {author} {\bibinfo {author} {\bibfnamefont {A.}~\bibnamefont
  {Roy}}\ and\ \bibinfo {author} {\bibfnamefont {M.}~\bibnamefont {Devoret}},\
  }\bibfield  {title} {\bibinfo {title} {Quantum-limited parametric
  amplification with josephson circuits in the regime of pump depletion},\
  }\href@noop {} {\bibfield  {journal} {\bibinfo  {journal} {Phys. Rev. B}\
  }\textbf {\bibinfo {volume} {98}},\ \bibinfo {pages} {045405} (\bibinfo
  {year} {2018})}\BibitemShut {NoStop}%
\bibitem [{\citenamefont {Eddins}\ \emph {et~al.}(2019)\citenamefont {Eddins},
  \citenamefont {Kreikebaum}, \citenamefont {Toyli}, \citenamefont
  {Levenson-Falk}, \citenamefont {Dove}, \citenamefont {Livingston},
  \citenamefont {Levitan}, \citenamefont {Govia}, \citenamefont {Clerk},\ and\
  \citenamefont {Siddiqi}}]{eddins2019highJPA}%
  \BibitemOpen
  \bibfield  {author} {\bibinfo {author} {\bibfnamefont {A.}~\bibnamefont
  {Eddins}}, \bibinfo {author} {\bibfnamefont {J.}~\bibnamefont {Kreikebaum}},
  \bibinfo {author} {\bibfnamefont {D.}~\bibnamefont {Toyli}}, \bibinfo
  {author} {\bibfnamefont {E.}~\bibnamefont {Levenson-Falk}}, \bibinfo {author}
  {\bibfnamefont {A.}~\bibnamefont {Dove}}, \bibinfo {author} {\bibfnamefont
  {W.}~\bibnamefont {Livingston}}, \bibinfo {author} {\bibfnamefont
  {B.}~\bibnamefont {Levitan}}, \bibinfo {author} {\bibfnamefont
  {L.}~\bibnamefont {Govia}}, \bibinfo {author} {\bibfnamefont
  {A.}~\bibnamefont {Clerk}},\ and\ \bibinfo {author} {\bibfnamefont
  {I.}~\bibnamefont {Siddiqi}},\ }\bibfield  {title} {\bibinfo {title}
  {High-efficiency measurement of an artificial atom embedded in a parametric
  amplifier},\ }\href@noop {} {\bibfield  {journal} {\bibinfo  {journal} {Phys.
  Rev. X}\ }\textbf {\bibinfo {volume} {9}},\ \bibinfo {pages} {011004}
  (\bibinfo {year} {2019})}\BibitemShut {NoStop}%
\bibitem [{\citenamefont {Yamamoto}\ \emph {et~al.}(2008)\citenamefont
  {Yamamoto}, \citenamefont {Inomata}, \citenamefont {Watanabe}, \citenamefont
  {Matsuba}, \citenamefont {Miyazaki}, \citenamefont {Oliver}, \citenamefont
  {Nakamura},\ and\ \citenamefont {Tsai}}]{yamamoto2008flux}%
  \BibitemOpen
  \bibfield  {author} {\bibinfo {author} {\bibfnamefont {T.}~\bibnamefont
  {Yamamoto}}, \bibinfo {author} {\bibfnamefont {K.}~\bibnamefont {Inomata}},
  \bibinfo {author} {\bibfnamefont {M.}~\bibnamefont {Watanabe}}, \bibinfo
  {author} {\bibfnamefont {K.}~\bibnamefont {Matsuba}}, \bibinfo {author}
  {\bibfnamefont {T.}~\bibnamefont {Miyazaki}}, \bibinfo {author}
  {\bibfnamefont {W.~D.}\ \bibnamefont {Oliver}}, \bibinfo {author}
  {\bibfnamefont {Y.}~\bibnamefont {Nakamura}},\ and\ \bibinfo {author}
  {\bibfnamefont {J.}~\bibnamefont {Tsai}},\ }\bibfield  {title} {\bibinfo
  {title} {Flux-driven josephson parametric amplifier},\ }\href@noop {}
  {\bibfield  {journal} {\bibinfo  {journal} {Appl. Phys. Lett.}\ }\textbf
  {\bibinfo {volume} {93}},\ \bibinfo {pages} {042510} (\bibinfo {year}
  {2008})}\BibitemShut {NoStop}%
\bibitem [{\citenamefont {Aumentado}(2020)}]{aumentado2020superconducting}%
  \BibitemOpen
  \bibfield  {author} {\bibinfo {author} {\bibfnamefont {J.}~\bibnamefont
  {Aumentado}},\ }\bibfield  {title} {\bibinfo {title} {Superconducting
  parametric amplifiers: The state of the art in josephson parametric
  amplifiers},\ }\href@noop {} {\bibfield  {journal} {\bibinfo  {journal} {IEEE
  Microw. Mag.}\ }\textbf {\bibinfo {volume} {21}},\ \bibinfo {pages} {45}
  (\bibinfo {year} {2020})}\BibitemShut {NoStop}%
\bibitem [{\citenamefont {Haus}\ and\ \citenamefont
  {Mullen}(1962)}]{haus1962quantum}%
  \BibitemOpen
  \bibfield  {author} {\bibinfo {author} {\bibfnamefont {H.~A.}\ \bibnamefont
  {Haus}}\ and\ \bibinfo {author} {\bibfnamefont {J.}~\bibnamefont {Mullen}},\
  }\bibfield  {title} {\bibinfo {title} {Quantum noise in linear amplifiers},\
  }\href@noop {} {\bibfield  {journal} {\bibinfo  {journal} {Phys. Rev.}\
  }\textbf {\bibinfo {volume} {128}},\ \bibinfo {pages} {2407} (\bibinfo {year}
  {1962})}\BibitemShut {NoStop}%
\bibitem [{\citenamefont {Sandberg}\ \emph {et~al.}(2008)\citenamefont
  {Sandberg}, \citenamefont {Wilson}, \citenamefont {Persson}, \citenamefont
  {Bauch}, \citenamefont {Johansson}, \citenamefont {Shumeiko}, \citenamefont
  {Duty},\ and\ \citenamefont {Delsing}}]{sandberg2008tuning}%
  \BibitemOpen
  \bibfield  {author} {\bibinfo {author} {\bibfnamefont {M.}~\bibnamefont
  {Sandberg}}, \bibinfo {author} {\bibfnamefont {C.}~\bibnamefont {Wilson}},
  \bibinfo {author} {\bibfnamefont {F.}~\bibnamefont {Persson}}, \bibinfo
  {author} {\bibfnamefont {T.}~\bibnamefont {Bauch}}, \bibinfo {author}
  {\bibfnamefont {G.}~\bibnamefont {Johansson}}, \bibinfo {author}
  {\bibfnamefont {V.}~\bibnamefont {Shumeiko}}, \bibinfo {author}
  {\bibfnamefont {T.}~\bibnamefont {Duty}},\ and\ \bibinfo {author}
  {\bibfnamefont {P.}~\bibnamefont {Delsing}},\ }\bibfield  {title} {\bibinfo
  {title} {Tuning the field in a microwave resonator faster than the photon
  lifetime},\ }\href@noop {} {\bibfield  {journal} {\bibinfo  {journal} {Appl.
  Phys. Lett.}\ }\textbf {\bibinfo {volume} {92}},\ \bibinfo {pages} {203501}
  (\bibinfo {year} {2008})}\BibitemShut {NoStop}%
\bibitem [{\citenamefont {Bergeal}\ \emph {et~al.}(2010)\citenamefont
  {Bergeal}, \citenamefont {Schackert}, \citenamefont {Metcalfe}, \citenamefont
  {Vijay}, \citenamefont {Manucharyan}, \citenamefont {Frunzio}, \citenamefont
  {Prober}, \citenamefont {Schoelkopf}, \citenamefont {Girvin},\ and\
  \citenamefont {Devoret}}]{vijay2010phase}%
  \BibitemOpen
  \bibfield  {author} {\bibinfo {author} {\bibfnamefont {N.}~\bibnamefont
  {Bergeal}}, \bibinfo {author} {\bibfnamefont {F.}~\bibnamefont {Schackert}},
  \bibinfo {author} {\bibfnamefont {M.}~\bibnamefont {Metcalfe}}, \bibinfo
  {author} {\bibfnamefont {R.}~\bibnamefont {Vijay}}, \bibinfo {author}
  {\bibfnamefont {V.}~\bibnamefont {Manucharyan}}, \bibinfo {author}
  {\bibfnamefont {L.}~\bibnamefont {Frunzio}}, \bibinfo {author} {\bibfnamefont
  {D.}~\bibnamefont {Prober}}, \bibinfo {author} {\bibfnamefont
  {R.}~\bibnamefont {Schoelkopf}}, \bibinfo {author} {\bibfnamefont
  {S.}~\bibnamefont {Girvin}},\ and\ \bibinfo {author} {\bibfnamefont
  {M.}~\bibnamefont {Devoret}},\ }\bibfield  {title} {\bibinfo {title}
  {Phase-preserving amplification near the quantum limit with a josephson ring
  modulator},\ }\href@noop {} {\bibfield  {journal} {\bibinfo  {journal}
  {Nature}\ }\textbf {\bibinfo {volume} {465}},\ \bibinfo {pages} {64}
  (\bibinfo {year} {2010})}\BibitemShut {NoStop}%
\bibitem [{\citenamefont {Nayfeh}\ and\ \citenamefont
  {Mook}(2008)}]{nayfeh2008nonlinear}%
  \BibitemOpen
  \bibfield  {author} {\bibinfo {author} {\bibfnamefont {A.~H.}\ \bibnamefont
  {Nayfeh}}\ and\ \bibinfo {author} {\bibfnamefont {D.~T.}\ \bibnamefont
  {Mook}},\ }\href@noop {} {\emph {\bibinfo {title} {Nonlinear oscillations}}}\
  (\bibinfo  {publisher} {John Wiley \& Sons},\ \bibinfo {year}
  {2008})\BibitemShut {NoStop}%
\bibitem [{\citenamefont {Lin}\ \emph {et~al.}(2014)\citenamefont {Lin},
  \citenamefont {Inomata}, \citenamefont {Koshino}, \citenamefont {Oliver},
  \citenamefont {Nakamura}, \citenamefont {Tsai},\ and\ \citenamefont
  {Yamamoto}}]{lin2014josephson}%
  \BibitemOpen
  \bibfield  {author} {\bibinfo {author} {\bibfnamefont {Z.}~\bibnamefont
  {Lin}}, \bibinfo {author} {\bibfnamefont {K.}~\bibnamefont {Inomata}},
  \bibinfo {author} {\bibfnamefont {K.}~\bibnamefont {Koshino}}, \bibinfo
  {author} {\bibfnamefont {W.}~\bibnamefont {Oliver}}, \bibinfo {author}
  {\bibfnamefont {Y.}~\bibnamefont {Nakamura}}, \bibinfo {author}
  {\bibfnamefont {J.-S.}\ \bibnamefont {Tsai}},\ and\ \bibinfo {author}
  {\bibfnamefont {T.}~\bibnamefont {Yamamoto}},\ }\bibfield  {title} {\bibinfo
  {title} {Josephson parametric phase-locked oscillator and its application to
  dispersive readout of superconducting qubits},\ }\href@noop {} {\bibfield
  {journal} {\bibinfo  {journal} {Nat. Commun.}\ }\textbf {\bibinfo {volume}
  {5}},\ \bibinfo {pages} {1} (\bibinfo {year} {2014})}\BibitemShut {NoStop}%
\bibitem [{\citenamefont {Krantz}\ \emph {et~al.}(2013)\citenamefont {Krantz},
  \citenamefont {Reshitnyk}, \citenamefont {Wustmann}, \citenamefont
  {Bylander}, \citenamefont {Gustavsson}, \citenamefont {Oliver}, \citenamefont
  {Duty}, \citenamefont {Shumeiko},\ and\ \citenamefont
  {Delsing}}]{krantz2013investigation}%
  \BibitemOpen
  \bibfield  {author} {\bibinfo {author} {\bibfnamefont {P.}~\bibnamefont
  {Krantz}}, \bibinfo {author} {\bibfnamefont {Y.}~\bibnamefont {Reshitnyk}},
  \bibinfo {author} {\bibfnamefont {W.}~\bibnamefont {Wustmann}}, \bibinfo
  {author} {\bibfnamefont {J.}~\bibnamefont {Bylander}}, \bibinfo {author}
  {\bibfnamefont {S.}~\bibnamefont {Gustavsson}}, \bibinfo {author}
  {\bibfnamefont {W.~D.}\ \bibnamefont {Oliver}}, \bibinfo {author}
  {\bibfnamefont {T.}~\bibnamefont {Duty}}, \bibinfo {author} {\bibfnamefont
  {V.}~\bibnamefont {Shumeiko}},\ and\ \bibinfo {author} {\bibfnamefont
  {P.}~\bibnamefont {Delsing}},\ }\bibfield  {title} {\bibinfo {title}
  {Investigation of nonlinear effects in josephson parametric oscillators used
  in circuit quantum electrodynamics},\ }\href@noop {} {\bibfield  {journal}
  {\bibinfo  {journal} {New J. Phys.}\ }\textbf {\bibinfo {volume} {15}},\
  \bibinfo {pages} {105002} (\bibinfo {year} {2013})}\BibitemShut {NoStop}%
\bibitem [{\citenamefont {Bengtsson}\ \emph {et~al.}(2018)\citenamefont
  {Bengtsson}, \citenamefont {Krantz}, \citenamefont {Simoen}, \citenamefont
  {Svensson}, \citenamefont {Schneider}, \citenamefont {Shumeiko},
  \citenamefont {Delsing},\ and\ \citenamefont
  {Bylander}}]{bengtsson2018nondegenerateJPO}%
  \BibitemOpen
  \bibfield  {author} {\bibinfo {author} {\bibfnamefont {A.}~\bibnamefont
  {Bengtsson}}, \bibinfo {author} {\bibfnamefont {P.}~\bibnamefont {Krantz}},
  \bibinfo {author} {\bibfnamefont {M.}~\bibnamefont {Simoen}}, \bibinfo
  {author} {\bibfnamefont {I.-M.}\ \bibnamefont {Svensson}}, \bibinfo {author}
  {\bibfnamefont {B.}~\bibnamefont {Schneider}}, \bibinfo {author}
  {\bibfnamefont {V.}~\bibnamefont {Shumeiko}}, \bibinfo {author}
  {\bibfnamefont {P.}~\bibnamefont {Delsing}},\ and\ \bibinfo {author}
  {\bibfnamefont {J.}~\bibnamefont {Bylander}},\ }\bibfield  {title} {\bibinfo
  {title} {Nondegenerate parametric oscillations in a tunable superconducting
  resonator},\ }\href@noop {} {\bibfield  {journal} {\bibinfo  {journal} {Phys.
  Rev. B}\ }\textbf {\bibinfo {volume} {97}},\ \bibinfo {pages} {144502}
  (\bibinfo {year} {2018})}\BibitemShut {NoStop}%
\bibitem [{\citenamefont {Meaney}\ \emph {et~al.}(2014)\citenamefont {Meaney},
  \citenamefont {Nha}, \citenamefont {Duty},\ and\ \citenamefont
  {Milburn}}]{squeezing_meaney2014quantum}%
  \BibitemOpen
  \bibfield  {author} {\bibinfo {author} {\bibfnamefont {C.~H.}\ \bibnamefont
  {Meaney}}, \bibinfo {author} {\bibfnamefont {H.}~\bibnamefont {Nha}},
  \bibinfo {author} {\bibfnamefont {T.}~\bibnamefont {Duty}},\ and\ \bibinfo
  {author} {\bibfnamefont {G.~J.}\ \bibnamefont {Milburn}},\ }\bibfield
  {title} {\bibinfo {title} {Quantum and classical nonlinear dynamics in a
  microwave cavity},\ }\href@noop {} {\bibfield  {journal} {\bibinfo  {journal}
  {Eur. Phys. J. Quantum Technol.}\ }\textbf {\bibinfo {volume} {1}},\ \bibinfo
  {pages} {1} (\bibinfo {year} {2014})}\BibitemShut {NoStop}%
\bibitem [{\citenamefont {Wustmann}\ and\ \citenamefont
  {Shumeiko}(2013)}]{squeezing2013parametric}%
  \BibitemOpen
  \bibfield  {author} {\bibinfo {author} {\bibfnamefont {W.}~\bibnamefont
  {Wustmann}}\ and\ \bibinfo {author} {\bibfnamefont {V.}~\bibnamefont
  {Shumeiko}},\ }\bibfield  {title} {\bibinfo {title} {Parametric resonance in
  tunable superconducting cavities},\ }\href@noop {} {\bibfield  {journal}
  {\bibinfo  {journal} {Phys. Rev. B}\ }\textbf {\bibinfo {volume} {87}},\
  \bibinfo {pages} {184501} (\bibinfo {year} {2013})}\BibitemShut {NoStop}%
\bibitem [{\citenamefont {Goto}(2016{\natexlab{a}})}]{goto2016bifurcation}%
  \BibitemOpen
  \bibfield  {author} {\bibinfo {author} {\bibfnamefont {H.}~\bibnamefont
  {Goto}},\ }\bibfield  {title} {\bibinfo {title} {Bifurcation-based adiabatic
  quantum computation with a nonlinear oscillator network},\ }\href@noop {}
  {\bibfield  {journal} {\bibinfo  {journal} {Sci. Rep.}\ }\textbf {\bibinfo
  {volume} {6}},\ \bibinfo {pages} {21686} (\bibinfo {year}
  {2016}{\natexlab{a}})}\BibitemShut {NoStop}%
\bibitem [{\citenamefont {Goto}(2016{\natexlab{b}})}]{goto2016universal}%
  \BibitemOpen
  \bibfield  {author} {\bibinfo {author} {\bibfnamefont {H.}~\bibnamefont
  {Goto}},\ }\bibfield  {title} {\bibinfo {title} {Universal quantum
  computation with a nonlinear oscillator network},\ }\href@noop {} {\bibfield
  {journal} {\bibinfo  {journal} {Phys. Rev. A}\ }\textbf {\bibinfo {volume}
  {93}},\ \bibinfo {pages} {050301} (\bibinfo {year}
  {2016}{\natexlab{b}})}\BibitemShut {NoStop}%
\bibitem [{\citenamefont {Puri}\ \emph {et~al.}(2017)\citenamefont {Puri},
  \citenamefont {Boutin},\ and\ \citenamefont {Blais}}]{puri2017engineering}%
  \BibitemOpen
  \bibfield  {author} {\bibinfo {author} {\bibfnamefont {S.}~\bibnamefont
  {Puri}}, \bibinfo {author} {\bibfnamefont {S.}~\bibnamefont {Boutin}},\ and\
  \bibinfo {author} {\bibfnamefont {A.}~\bibnamefont {Blais}},\ }\bibfield
  {title} {\bibinfo {title} {Engineering the quantum states of light in a
  kerr-nonlinear resonator by two-photon driving},\ }\href@noop {} {\bibfield
  {journal} {\bibinfo  {journal} {npj Quantum Inf.}\ }\textbf {\bibinfo
  {volume} {3}},\ \bibinfo {pages} {1} (\bibinfo {year} {2017})}\BibitemShut
  {NoStop}%
\bibitem [{\citenamefont {Mirrahimi}\ \emph {et~al.}(2014)\citenamefont
  {Mirrahimi}, \citenamefont {Leghtas}, \citenamefont {Albert}, \citenamefont
  {Touzard}, \citenamefont {Schoelkopf}, \citenamefont {Jiang},\ and\
  \citenamefont {Devoret}}]{catqubit2014dynamically}%
  \BibitemOpen
  \bibfield  {author} {\bibinfo {author} {\bibfnamefont {M.}~\bibnamefont
  {Mirrahimi}}, \bibinfo {author} {\bibfnamefont {Z.}~\bibnamefont {Leghtas}},
  \bibinfo {author} {\bibfnamefont {V.~V.}\ \bibnamefont {Albert}}, \bibinfo
  {author} {\bibfnamefont {S.}~\bibnamefont {Touzard}}, \bibinfo {author}
  {\bibfnamefont {R.~J.}\ \bibnamefont {Schoelkopf}}, \bibinfo {author}
  {\bibfnamefont {L.}~\bibnamefont {Jiang}},\ and\ \bibinfo {author}
  {\bibfnamefont {M.~H.}\ \bibnamefont {Devoret}},\ }\bibfield  {title}
  {\bibinfo {title} {Dynamically protected cat-qubits: a new paradigm for
  universal quantum computation},\ }\href@noop {} {\bibfield  {journal}
  {\bibinfo  {journal} {New J. Phys.}\ }\textbf {\bibinfo {volume} {16}},\
  \bibinfo {pages} {045014} (\bibinfo {year} {2014})}\BibitemShut {NoStop}%
\bibitem [{\citenamefont {Krantz}\ \emph {et~al.}(2016)\citenamefont {Krantz},
  \citenamefont {Bengtsson}, \citenamefont {Simoen}, \citenamefont
  {Gustavsson}, \citenamefont {Shumeiko}, \citenamefont {Oliver}, \citenamefont
  {Wilson}, \citenamefont {Delsing},\ and\ \citenamefont
  {Bylander}}]{singleshot_krantz}%
  \BibitemOpen
  \bibfield  {author} {\bibinfo {author} {\bibfnamefont {P.}~\bibnamefont
  {Krantz}}, \bibinfo {author} {\bibfnamefont {A.}~\bibnamefont {Bengtsson}},
  \bibinfo {author} {\bibfnamefont {M.}~\bibnamefont {Simoen}}, \bibinfo
  {author} {\bibfnamefont {S.}~\bibnamefont {Gustavsson}}, \bibinfo {author}
  {\bibfnamefont {V.}~\bibnamefont {Shumeiko}}, \bibinfo {author}
  {\bibfnamefont {W.}~\bibnamefont {Oliver}}, \bibinfo {author} {\bibfnamefont
  {C.}~\bibnamefont {Wilson}}, \bibinfo {author} {\bibfnamefont
  {P.}~\bibnamefont {Delsing}},\ and\ \bibinfo {author} {\bibfnamefont
  {J.}~\bibnamefont {Bylander}},\ }\bibfield  {title} {\bibinfo {title}
  {Single-shot read-out of a superconducting qubit using a josephson parametric
  oscillator},\ }\href@noop {} {\bibfield  {journal} {\bibinfo  {journal} {Nat.
  Commun.}\ }\textbf {\bibinfo {volume} {7}},\ \bibinfo {pages} {11417}
  (\bibinfo {year} {2016})}\BibitemShut {NoStop}%
\bibitem [{\citenamefont {Wilson}\ \emph {et~al.}(2010)\citenamefont {Wilson},
  \citenamefont {Duty}, \citenamefont {Sandberg}, \citenamefont {Persson},
  \citenamefont {Shumeiko},\ and\ \citenamefont {Delsing}}]{Casimir}%
  \BibitemOpen
  \bibfield  {author} {\bibinfo {author} {\bibfnamefont {C.}~\bibnamefont
  {Wilson}}, \bibinfo {author} {\bibfnamefont {T.}~\bibnamefont {Duty}},
  \bibinfo {author} {\bibfnamefont {M.}~\bibnamefont {Sandberg}}, \bibinfo
  {author} {\bibfnamefont {F.}~\bibnamefont {Persson}}, \bibinfo {author}
  {\bibfnamefont {V.}~\bibnamefont {Shumeiko}},\ and\ \bibinfo {author}
  {\bibfnamefont {P.}~\bibnamefont {Delsing}},\ }\bibfield  {title} {\bibinfo
  {title} {Photon generation in an electromagnetic cavity with a time-dependent
  boundary},\ }\href@noop {} {\bibfield  {journal} {\bibinfo  {journal} {Phys.
  Rev. Lett.}\ }\textbf {\bibinfo {volume} {105}},\ \bibinfo {pages} {233907}
  (\bibinfo {year} {2010})}\BibitemShut {NoStop}%
\bibitem [{\citenamefont {Yurke}\ \emph {et~al.}(1991)\citenamefont {Yurke},
  \citenamefont {Movshovich}, \citenamefont {Kaminsky}, \citenamefont {Bryant},
  \citenamefont {Smith}, \citenamefont {Silver},\ and\ \citenamefont
  {Simon}}]{yurke_noise1991behavior}%
  \BibitemOpen
  \bibfield  {author} {\bibinfo {author} {\bibfnamefont {B.}~\bibnamefont
  {Yurke}}, \bibinfo {author} {\bibfnamefont {R.}~\bibnamefont {Movshovich}},
  \bibinfo {author} {\bibfnamefont {P.}~\bibnamefont {Kaminsky}}, \bibinfo
  {author} {\bibfnamefont {P.}~\bibnamefont {Bryant}}, \bibinfo {author}
  {\bibfnamefont {A.}~\bibnamefont {Smith}}, \bibinfo {author} {\bibfnamefont
  {A.}~\bibnamefont {Silver}},\ and\ \bibinfo {author} {\bibfnamefont
  {R.}~\bibnamefont {Simon}},\ }\bibfield  {title} {\bibinfo {title} {Behavior
  of noise in a nondegenerate josephson-parametric amplifier},\ }\href@noop {}
  {\bibfield  {journal} {\bibinfo  {journal} {IEEE Trans. Magn.}\ }\textbf
  {\bibinfo {volume} {27}},\ \bibinfo {pages} {3374} (\bibinfo {year}
  {1991})}\BibitemShut {NoStop}%
\bibitem [{\citenamefont {Reid}\ and\ \citenamefont {Yurke}(1992)}]{JPO_Yurke}%
  \BibitemOpen
  \bibfield  {author} {\bibinfo {author} {\bibfnamefont {M.}~\bibnamefont
  {Reid}}\ and\ \bibinfo {author} {\bibfnamefont {B.}~\bibnamefont {Yurke}},\
  }\bibfield  {title} {\bibinfo {title} {Effect of bistability and
  superpositions on quantum statistics in degenerate parametric oscillation},\
  }\href@noop {} {\bibfield  {journal} {\bibinfo  {journal} {Phys. Rev. A}\
  }\textbf {\bibinfo {volume} {46}},\ \bibinfo {pages} {4131} (\bibinfo {year}
  {1992})}\BibitemShut {NoStop}%
\bibitem [{\citenamefont {Olsson}\ and\ \citenamefont
  {Claeson}(1988)}]{olsson1988lowJPO}%
  \BibitemOpen
  \bibfield  {author} {\bibinfo {author} {\bibfnamefont {H.}~\bibnamefont
  {Olsson}}\ and\ \bibinfo {author} {\bibfnamefont {T.}~\bibnamefont
  {Claeson}},\ }\bibfield  {title} {\bibinfo {title} {Low-noise josephson
  parametric amplification and oscillations at 9 ghz},\ }\href@noop {}
  {\bibfield  {journal} {\bibinfo  {journal} {J. Appl. Phys.}\ }\textbf
  {\bibinfo {volume} {64}},\ \bibinfo {pages} {5234} (\bibinfo {year}
  {1988})}\BibitemShut {NoStop}%
\bibitem [{\citenamefont {Hajimiri}\ and\ \citenamefont
  {Lee}(1998)}]{uni_hajimiri1998general}%
  \BibitemOpen
  \bibfield  {author} {\bibinfo {author} {\bibfnamefont {A.}~\bibnamefont
  {Hajimiri}}\ and\ \bibinfo {author} {\bibfnamefont {T.~H.}\ \bibnamefont
  {Lee}},\ }\bibfield  {title} {\bibinfo {title} {A general theory of phase
  noise in electrical oscillators},\ }\href@noop {} {\bibfield  {journal}
  {\bibinfo  {journal} {IEEE J. Solid-State Circuits}\ }\textbf {\bibinfo
  {volume} {33}},\ \bibinfo {pages} {179} (\bibinfo {year} {1998})}\BibitemShut
  {NoStop}%
\bibitem [{\citenamefont {Demir}\ \emph {et~al.}(2000)\citenamefont {Demir},
  \citenamefont {Mehrotra},\ and\ \citenamefont
  {Roychowdhury}}]{demir2000phaseunifying}%
  \BibitemOpen
  \bibfield  {author} {\bibinfo {author} {\bibfnamefont {A.}~\bibnamefont
  {Demir}}, \bibinfo {author} {\bibfnamefont {A.}~\bibnamefont {Mehrotra}},\
  and\ \bibinfo {author} {\bibfnamefont {J.}~\bibnamefont {Roychowdhury}},\
  }\bibfield  {title} {\bibinfo {title} {Phase noise in oscillators: A unifying
  theory and numerical methods for characterization},\ }\href@noop {}
  {\bibfield  {journal} {\bibinfo  {journal} {IEEE Trans. Circuits Syst.
  Fundam. Theory Appl.}\ }\textbf {\bibinfo {volume} {47}},\ \bibinfo {pages}
  {655} (\bibinfo {year} {2000})}\BibitemShut {NoStop}%
\bibitem [{\citenamefont {Sauvage}(1977)}]{uni_sauvage1977phase}%
  \BibitemOpen
  \bibfield  {author} {\bibinfo {author} {\bibfnamefont {G.}~\bibnamefont
  {Sauvage}},\ }\bibfield  {title} {\bibinfo {title} {Phase noise in
  oscillators: A mathematical analysis of leeson's model},\ }\href@noop {}
  {\bibfield  {journal} {\bibinfo  {journal} {IEEE Trans. Instrum. Meas}\
  }\textbf {\bibinfo {volume} {26}},\ \bibinfo {pages} {408} (\bibinfo {year}
  {1977})}\BibitemShut {NoStop}%
\bibitem [{\citenamefont {Reynaud}\ \emph {et~al.}(1987)\citenamefont
  {Reynaud}, \citenamefont {Fabre},\ and\ \citenamefont
  {Giacobino}}]{opo_reynaud1987quantum}%
  \BibitemOpen
  \bibfield  {author} {\bibinfo {author} {\bibfnamefont {S.}~\bibnamefont
  {Reynaud}}, \bibinfo {author} {\bibfnamefont {C.}~\bibnamefont {Fabre}},\
  and\ \bibinfo {author} {\bibfnamefont {E.}~\bibnamefont {Giacobino}},\
  }\bibfield  {title} {\bibinfo {title} {Quantum fluctuations in a two-mode
  parametric oscillator},\ }\href@noop {} {\bibfield  {journal} {\bibinfo
  {journal} {JOSA B}\ }\textbf {\bibinfo {volume} {4}},\ \bibinfo {pages}
  {1520} (\bibinfo {year} {1987})}\BibitemShut {NoStop}%
\bibitem [{\citenamefont {Courtois}\ \emph {et~al.}(1991)\citenamefont
  {Courtois}, \citenamefont {Smith}, \citenamefont {Fabre},\ and\ \citenamefont
  {Reynaud}}]{imp_opo_1991phase}%
  \BibitemOpen
  \bibfield  {author} {\bibinfo {author} {\bibfnamefont {J.}~\bibnamefont
  {Courtois}}, \bibinfo {author} {\bibfnamefont {A.}~\bibnamefont {Smith}},
  \bibinfo {author} {\bibfnamefont {C.}~\bibnamefont {Fabre}},\ and\ \bibinfo
  {author} {\bibfnamefont {S.}~\bibnamefont {Reynaud}},\ }\bibfield  {title}
  {\bibinfo {title} {Phase diffusion and quantum noise in the optical
  parametric oscillator: a semiclassical approach},\ }\href@noop {} {\bibfield
  {journal} {\bibinfo  {journal} {J. Mod. Opt.}\ }\textbf {\bibinfo {volume}
  {38}},\ \bibinfo {pages} {177} (\bibinfo {year} {1991})}\BibitemShut
  {NoStop}%
\bibitem [{\citenamefont {Henry}(1983{\natexlab{a}})}]{laser_henry1983theory}%
  \BibitemOpen
  \bibfield  {author} {\bibinfo {author} {\bibfnamefont {C.}~\bibnamefont
  {Henry}},\ }\bibfield  {title} {\bibinfo {title} {Theory of the phase noise
  and power spectrum of a single mode injection laser},\ }\href@noop {}
  {\bibfield  {journal} {\bibinfo  {journal} {IEEE J. Quantum Electron.}\
  }\textbf {\bibinfo {volume} {19}},\ \bibinfo {pages} {1391} (\bibinfo {year}
  {1983}{\natexlab{a}})}\BibitemShut {NoStop}%
\bibitem [{\citenamefont {Debuisschert}\ \emph {et~al.}(1989)\citenamefont
  {Debuisschert}, \citenamefont {Reynaud}, \citenamefont {Heidmann},
  \citenamefont {Giacobino},\ and\ \citenamefont
  {Fabre}}]{opo_debuisschert1989observation}%
  \BibitemOpen
  \bibfield  {author} {\bibinfo {author} {\bibfnamefont {T.}~\bibnamefont
  {Debuisschert}}, \bibinfo {author} {\bibfnamefont {S.}~\bibnamefont
  {Reynaud}}, \bibinfo {author} {\bibfnamefont {A.}~\bibnamefont {Heidmann}},
  \bibinfo {author} {\bibfnamefont {E.}~\bibnamefont {Giacobino}},\ and\
  \bibinfo {author} {\bibfnamefont {C.}~\bibnamefont {Fabre}},\ }\bibfield
  {title} {\bibinfo {title} {Observation of large quantum noise reduction using
  an optical parametric oscillator},\ }\href@noop {} {\bibfield  {journal}
  {\bibinfo  {journal} {Quantum Opt.}\ }\textbf {\bibinfo {volume} {1}},\
  \bibinfo {pages} {3} (\bibinfo {year} {1989})}\BibitemShut {NoStop}%
\bibitem [{\citenamefont {Lin}\ \emph {et~al.}(2015)\citenamefont {Lin},
  \citenamefont {Nakamura},\ and\ \citenamefont {Dykman}}]{dykman2015critical}%
  \BibitemOpen
  \bibfield  {author} {\bibinfo {author} {\bibfnamefont {Z.}~\bibnamefont
  {Lin}}, \bibinfo {author} {\bibfnamefont {Y.}~\bibnamefont {Nakamura}},\ and\
  \bibinfo {author} {\bibfnamefont {M.}~\bibnamefont {Dykman}},\ }\bibfield
  {title} {\bibinfo {title} {Critical fluctuations and the rates of interstate
  switching near the excitation threshold of a quantum parametric oscillator},\
  }\href@noop {} {\bibfield  {journal} {\bibinfo  {journal} {Phys. Rev. E}\
  }\textbf {\bibinfo {volume} {92}},\ \bibinfo {pages} {022105} (\bibinfo
  {year} {2015})}\BibitemShut {NoStop}%
\bibitem [{\citenamefont {Hinkley}\ and\ \citenamefont
  {Freed}(1969)}]{hinkley1969directlaser}%
  \BibitemOpen
  \bibfield  {author} {\bibinfo {author} {\bibfnamefont {E.}~\bibnamefont
  {Hinkley}}\ and\ \bibinfo {author} {\bibfnamefont {C.}~\bibnamefont
  {Freed}},\ }\bibfield  {title} {\bibinfo {title} {Direct observation of the
  lorentzian line shape as limited by quantum phase noise in a laser above
  threshold},\ }\href@noop {} {\bibfield  {journal} {\bibinfo  {journal} {Phys.
  Rev. Lett.}\ }\textbf {\bibinfo {volume} {23}},\ \bibinfo {pages} {277}
  (\bibinfo {year} {1969})}\BibitemShut {NoStop}%
\bibitem [{\citenamefont {Henry}(1983{\natexlab{b}})}]{henry1983theoryST}%
  \BibitemOpen
  \bibfield  {author} {\bibinfo {author} {\bibfnamefont {C.}~\bibnamefont
  {Henry}},\ }\bibfield  {title} {\bibinfo {title} {Theory of the phase noise
  and power spectrum of a single mode injection laser},\ }\href@noop {}
  {\bibfield  {journal} {\bibinfo  {journal} {IEEE J. Quantum Electron.}\
  }\textbf {\bibinfo {volume} {19}},\ \bibinfo {pages} {1391} (\bibinfo {year}
  {1983}{\natexlab{b}})}\BibitemShut {NoStop}%
\bibitem [{\citenamefont {Frunzio}\ \emph {et~al.}(2005)\citenamefont
  {Frunzio}, \citenamefont {Wallraff}, \citenamefont {Schuster}, \citenamefont
  {Majer},\ and\ \citenamefont {Schoelkopf}}]{frunzio2005fabrication}%
  \BibitemOpen
  \bibfield  {author} {\bibinfo {author} {\bibfnamefont {L.}~\bibnamefont
  {Frunzio}}, \bibinfo {author} {\bibfnamefont {A.}~\bibnamefont {Wallraff}},
  \bibinfo {author} {\bibfnamefont {D.}~\bibnamefont {Schuster}}, \bibinfo
  {author} {\bibfnamefont {J.}~\bibnamefont {Majer}},\ and\ \bibinfo {author}
  {\bibfnamefont {R.}~\bibnamefont {Schoelkopf}},\ }\bibfield  {title}
  {\bibinfo {title} {Fabrication and characterization of superconducting
  circuit qed devices for quantum computation},\ }\href@noop {} {\bibfield
  {journal} {\bibinfo  {journal} {IEEE Trans. Appl. Supercond.}\ }\textbf
  {\bibinfo {volume} {15}},\ \bibinfo {pages} {860} (\bibinfo {year}
  {2005})}\BibitemShut {NoStop}%
\bibitem [{\citenamefont {Haken}(1975)}]{RevModPhys_equilibrium}%
  \BibitemOpen
  \bibfield  {author} {\bibinfo {author} {\bibfnamefont {H.}~\bibnamefont
  {Haken}},\ }\bibfield  {title} {\bibinfo {title} {Cooperative phenomena in
  systems far from thermal equilibrium and in nonphysical systems},\ }\href
  {https://doi.org/10.1103/RevModPhys.47.67} {\bibfield  {journal} {\bibinfo
  {journal} {Rev. Mod. Phys.}\ }\textbf {\bibinfo {volume} {47}},\ \bibinfo
  {pages} {67} (\bibinfo {year} {1975})}\BibitemShut {NoStop}%
\bibitem [{\citenamefont {Wustmann}\ and\ \citenamefont
  {Shumeiko}(2019)}]{wustmann2019_review_parametric}%
  \BibitemOpen
  \bibfield  {author} {\bibinfo {author} {\bibfnamefont {W.}~\bibnamefont
  {Wustmann}}\ and\ \bibinfo {author} {\bibfnamefont {V.}~\bibnamefont
  {Shumeiko}},\ }\bibfield  {title} {\bibinfo {title} {Parametric effects in
  circuit quantum electrodynamics},\ }\href@noop {} {\bibfield  {journal}
  {\bibinfo  {journal} {Low Temp. Phys.}\ }\textbf {\bibinfo {volume} {45}},\
  \bibinfo {pages} {848} (\bibinfo {year} {2019})}\BibitemShut {NoStop}%
\bibitem [{\citenamefont {Dykman}(2012)}]{dykman2012fluctuating}%
  \BibitemOpen
  \bibfield  {author} {\bibinfo {author} {\bibfnamefont {M.}~\bibnamefont
  {Dykman}},\ }\href@noop {} {\emph {\bibinfo {title} {Fluctuating nonlinear
  oscillators: from nanomechanics to quantum superconducting circuits}}}\
  (\bibinfo  {publisher} {Oxford University Press},\ \bibinfo {year}
  {2012})\BibitemShut {NoStop}%
\bibitem [{\citenamefont {Gao}\ \emph {et~al.}(2007)\citenamefont {Gao},
  \citenamefont {Zmuidzinas}, \citenamefont {Mazin}, \citenamefont {LeDuc},\
  and\ \citenamefont {Day}}]{gao2007noise}%
  \BibitemOpen
  \bibfield  {author} {\bibinfo {author} {\bibfnamefont {J.}~\bibnamefont
  {Gao}}, \bibinfo {author} {\bibfnamefont {J.}~\bibnamefont {Zmuidzinas}},
  \bibinfo {author} {\bibfnamefont {B.~A.}\ \bibnamefont {Mazin}}, \bibinfo
  {author} {\bibfnamefont {H.~G.}\ \bibnamefont {LeDuc}},\ and\ \bibinfo
  {author} {\bibfnamefont {P.~K.}\ \bibnamefont {Day}},\ }\bibfield  {title}
  {\bibinfo {title} {Noise properties of superconducting coplanar waveguide
  microwave resonators},\ }\href@noop {} {\bibfield  {journal} {\bibinfo
  {journal} {Appl. Phys. Lett.}\ }\textbf {\bibinfo {volume} {90}},\ \bibinfo
  {pages} {102507} (\bibinfo {year} {2007})}\BibitemShut {NoStop}%
\bibitem [{\citenamefont {Gao}\ \emph {et~al.}(2011)\citenamefont {Gao},
  \citenamefont {Vale}, \citenamefont {Mates}, \citenamefont {Schmidt},
  \citenamefont {Hilton}, \citenamefont {Irwin}, \citenamefont {Mallet},
  \citenamefont {Castellanos-Beltran}, \citenamefont {Lehnert}, \citenamefont
  {Zmuidzinas} \emph {et~al.}}]{gao2011vacuum}%
  \BibitemOpen
  \bibfield  {author} {\bibinfo {author} {\bibfnamefont {J.}~\bibnamefont
  {Gao}}, \bibinfo {author} {\bibfnamefont {L.}~\bibnamefont {Vale}}, \bibinfo
  {author} {\bibfnamefont {J.}~\bibnamefont {Mates}}, \bibinfo {author}
  {\bibfnamefont {D.}~\bibnamefont {Schmidt}}, \bibinfo {author} {\bibfnamefont
  {G.}~\bibnamefont {Hilton}}, \bibinfo {author} {\bibfnamefont
  {K.}~\bibnamefont {Irwin}}, \bibinfo {author} {\bibfnamefont
  {F.}~\bibnamefont {Mallet}}, \bibinfo {author} {\bibfnamefont
  {M.}~\bibnamefont {Castellanos-Beltran}}, \bibinfo {author} {\bibfnamefont
  {K.}~\bibnamefont {Lehnert}}, \bibinfo {author} {\bibfnamefont
  {J.}~\bibnamefont {Zmuidzinas}}, \emph {et~al.},\ }\bibfield  {title}
  {\bibinfo {title} {Strongly quadrature-dependent noise in superconducting
  microresonators measured at the vacuum-noise limit},\ }\href@noop {}
  {\bibfield  {journal} {\bibinfo  {journal} {Appl. Phys. Lett.}\ }\textbf
  {\bibinfo {volume} {98}},\ \bibinfo {pages} {232508} (\bibinfo {year}
  {2011})}\BibitemShut {NoStop}%
\bibitem [{\citenamefont {Welch}(1967)}]{welch1967use}%
  \BibitemOpen
  \bibfield  {author} {\bibinfo {author} {\bibfnamefont {P.}~\bibnamefont
  {Welch}},\ }\bibfield  {title} {\bibinfo {title} {The use of fast fourier
  transform for the estimation of power spectra: a method based on time
  averaging over short, modified periodograms},\ }\href@noop {} {\bibfield
  {journal} {\bibinfo  {journal} {IEEE Trans. Audio Electroacoust.}\ }\textbf
  {\bibinfo {volume} {15}},\ \bibinfo {pages} {70} (\bibinfo {year}
  {1967})}\BibitemShut {NoStop}%
\bibitem [{\citenamefont {Dugan}\ \emph {et~al.}(2002)\citenamefont {Dugan},
  \citenamefont {McGranaghan},\ and\ \citenamefont
  {Beaty}}]{dugan1996electrical}%
  \BibitemOpen
  \bibfield  {author} {\bibinfo {author} {\bibfnamefont {R.~C.}\ \bibnamefont
  {Dugan}}, \bibinfo {author} {\bibfnamefont {M.~F.}\ \bibnamefont
  {McGranaghan}},\ and\ \bibinfo {author} {\bibfnamefont {H.~W.}\ \bibnamefont
  {Beaty}},\ }\href@noop {} {\emph {\bibinfo {title} {Electrical Power Systems
  Quality}}}\ (\bibinfo  {publisher} {McGraw-Hill},\ \bibinfo {year}
  {2002})\BibitemShut {NoStop}%
\bibitem [{\citenamefont {Tahir}\ and\ \citenamefont
  {Mazumder}(2014)}]{filter_2014improving}%
  \BibitemOpen
  \bibfield  {author} {\bibinfo {author} {\bibfnamefont {M.}~\bibnamefont
  {Tahir}}\ and\ \bibinfo {author} {\bibfnamefont {S.~K.}\ \bibnamefont
  {Mazumder}},\ }\bibfield  {title} {\bibinfo {title} {Improving dynamic
  response of active harmonic compensator using digital comb filter},\
  }\href@noop {} {\bibfield  {journal} {\bibinfo  {journal} {IEEE Trans. Emerg.
  Sel. Topics Power Electron.}\ }\textbf {\bibinfo {volume} {2}},\ \bibinfo
  {pages} {994} (\bibinfo {year} {2014})}\BibitemShut {NoStop}%
\bibitem [{\citenamefont {Machlup}(1954)}]{RTS_lorentzian_noise}%
  \BibitemOpen
  \bibfield  {author} {\bibinfo {author} {\bibfnamefont {S.}~\bibnamefont
  {Machlup}},\ }\bibfield  {title} {\bibinfo {title} {Noise in semiconductors:
  spectrum of a two-parameter random signal},\ }\href@noop {} {\bibfield
  {journal} {\bibinfo  {journal} {J. Appl. Phys.}\ }\textbf {\bibinfo {volume}
  {25}},\ \bibinfo {pages} {341} (\bibinfo {year} {1954})}\BibitemShut
  {NoStop}%
\bibitem [{\citenamefont {Pan}\ \emph {et~al.}(2022)\citenamefont {Pan},
  \citenamefont {Yuan}, \citenamefont {Zhou}, \citenamefont {Zhang},
  \citenamefont {Li}, \citenamefont {Liu}, \citenamefont {Jiang}, \citenamefont
  {Catelani}, \citenamefont {Hu},\ and\ \citenamefont
  {Yan}}]{feyyan_engineering}%
  \BibitemOpen
  \bibfield  {author} {\bibinfo {author} {\bibfnamefont {X.}~\bibnamefont
  {Pan}}, \bibinfo {author} {\bibfnamefont {H.}~\bibnamefont {Yuan}}, \bibinfo
  {author} {\bibfnamefont {Y.}~\bibnamefont {Zhou}}, \bibinfo {author}
  {\bibfnamefont {L.}~\bibnamefont {Zhang}}, \bibinfo {author} {\bibfnamefont
  {J.}~\bibnamefont {Li}}, \bibinfo {author} {\bibfnamefont {S.}~\bibnamefont
  {Liu}}, \bibinfo {author} {\bibfnamefont {Z.~H.}\ \bibnamefont {Jiang}},
  \bibinfo {author} {\bibfnamefont {G.}~\bibnamefont {Catelani}}, \bibinfo
  {author} {\bibfnamefont {L.}~\bibnamefont {Hu}},\ and\ \bibinfo {author}
  {\bibfnamefont {F.}~\bibnamefont {Yan}},\ }\bibfield  {title} {\bibinfo
  {title} {Engineering superconducting qubits to reduce quasiparticles and
  charge noise},\ }\href@noop {} {\bibfield  {journal} {\bibinfo  {journal}
  {arXiv:2202.01435}\ } (\bibinfo {year} {2022})}\BibitemShut {NoStop}%
\bibitem [{\citenamefont {Friedman}\ and\ \citenamefont
  {Han}(2003)}]{friedman2003classical}%
  \BibitemOpen
  \bibfield  {author} {\bibinfo {author} {\bibfnamefont {J.~R.}\ \bibnamefont
  {Friedman}}\ and\ \bibinfo {author} {\bibfnamefont {S.}~\bibnamefont {Han}},\
  }\href@noop {} {\emph {\bibinfo {title} {Exploring the quantum/classical
  frontier: recent advances in macroscopic quantum phenomena}}}\ (\bibinfo
  {publisher} {Nova Publishers},\ \bibinfo {year} {2003})\BibitemShut {NoStop}%
\bibitem [{\citenamefont {Razavy}(2013)}]{razavy2013quantum_tunnelling}%
  \BibitemOpen
  \bibfield  {author} {\bibinfo {author} {\bibfnamefont {M.}~\bibnamefont
  {Razavy}},\ }\href@noop {} {\emph {\bibinfo {title} {Quantum theory of
  tunneling}}}\ (\bibinfo  {publisher} {World Scientific},\ \bibinfo {year}
  {2013})\BibitemShut {NoStop}%
\bibitem [{\citenamefont {Dykman}(2007)}]{dykman_quantum2007critical}%
  \BibitemOpen
  \bibfield  {author} {\bibinfo {author} {\bibfnamefont {M.}~\bibnamefont
  {Dykman}},\ }\bibfield  {title} {\bibinfo {title} {Critical exponents in
  metastable decay via quantum activation},\ }\href@noop {} {\bibfield
  {journal} {\bibinfo  {journal} {Phys. Rev. E}\ }\textbf {\bibinfo {volume}
  {75}},\ \bibinfo {pages} {011101} (\bibinfo {year} {2007})}\BibitemShut
  {NoStop}%
\bibitem [{\citenamefont {Devoret}\ \emph {et~al.}(1985)\citenamefont
  {Devoret}, \citenamefont {Martinis},\ and\ \citenamefont
  {Clarke}}]{devoret_tunnelling_temperature}%
  \BibitemOpen
  \bibfield  {author} {\bibinfo {author} {\bibfnamefont {M.~H.}\ \bibnamefont
  {Devoret}}, \bibinfo {author} {\bibfnamefont {J.~M.}\ \bibnamefont
  {Martinis}},\ and\ \bibinfo {author} {\bibfnamefont {J.}~\bibnamefont
  {Clarke}},\ }\bibfield  {title} {\bibinfo {title} {Measurements of
  macroscopic quantum tunneling out of the zero-voltage state of a
  current-biased josephson junction},\ }\href@noop {} {\bibfield  {journal}
  {\bibinfo  {journal} {Physical review letters}\ }\textbf {\bibinfo {volume}
  {55}},\ \bibinfo {pages} {1908} (\bibinfo {year} {1985})}\BibitemShut
  {NoStop}%
\bibitem [{\citenamefont {Markovi{\'c}}\ \emph {et~al.}(2019)\citenamefont
  {Markovi{\'c}}, \citenamefont {Pillet}, \citenamefont {Flurin}, \citenamefont
  {Roch},\ and\ \citenamefont {Huard}}]{markovic2019injection}%
  \BibitemOpen
  \bibfield  {author} {\bibinfo {author} {\bibfnamefont {D.}~\bibnamefont
  {Markovi{\'c}}}, \bibinfo {author} {\bibfnamefont {J.-D.}\ \bibnamefont
  {Pillet}}, \bibinfo {author} {\bibfnamefont {E.}~\bibnamefont {Flurin}},
  \bibinfo {author} {\bibfnamefont {N.}~\bibnamefont {Roch}},\ and\ \bibinfo
  {author} {\bibfnamefont {B.}~\bibnamefont {Huard}},\ }\bibfield  {title}
  {\bibinfo {title} {Injection locking and parametric locking in a
  superconducting circuit},\ }\href@noop {} {\bibfield  {journal} {\bibinfo
  {journal} {Phys. Rev. Applied}\ }\textbf {\bibinfo {volume} {12}},\ \bibinfo
  {pages} {024034} (\bibinfo {year} {2019})}\BibitemShut {NoStop}%
\end{thebibliography}%

\end{document}